\documentclass[aps,prd,amsmath,amssymb,reprint,nofootinbib,longbibliography]{revtex4-2}
\usepackage{graphicx}       % Include figure files
\usepackage{physics}	    % Physics
\usepackage[hidelinks]{hyperref}
\usepackage[nameinlink,capitalise]{cleveref}
\renewcommand*{\vec}[1]{\vb{#1}}
\newcommand*{\ue}{\mathrm{e}}
\newcommand*{\ui}{\mathrm{i}}
\graphicspath{{figures/}}

\begin{document}
\preprint{KCL-PC-TH-2019-18}
\preprint{arXiv:1903.09690}
\title{Evanescent Gravitational Waves}
\author{Sebastian Golat}    \email{sebastian.golat@mensa.cz}
\author{Eugene A. Lim}      \email{eugene.a.lim@gmail.com}
\author{Francisco J. Rodr\'iguez-Fortu\~no} \email{francisco.rodriguez\_fortuno@kcl.ac.uk}
\affiliation{%
Department of Physics, King’s College London, Strand, London WC2R 2LS, United Kingdom}
\date{\today}
\begin{abstract}
    We describe the properties of evanescent gravitational waves (EGWs)---wave solutions of Einstein equations which decay exponentially in some direction while propagating in another. Evanescent waves are well-known in acoustics and optics and have recently received much attention due to their extraordinary properties such as their transverse spin and spin-momentum locking. We show that EGWs possess similarly remarkable properties, carrying transverse spin angular momenta and driving freely falling test masses along in elliptical trajectories. Hence, test masses on a plane transverse to the direction of propagation exhibit correlated vector and scalar-like deformation---correlations which can be used to distinguish it from modified gravity. We demonstrate that EGWs are present and dominant in the vicinity of sub-wavelength sources such as orbiting binaries.
\end{abstract}
\maketitle

\section{Introduction}
    {Evanescent waves, or fields, are solutions to the wave equation which instead of propagating away from the source, decay exponentially. While evanescent waves have been known for a very long time, only recently have they been intensively studied, following the increased interest in small scale physics. In nanophotonics, evanescent waves play a dominant role \cite{Fornel2001}. Recent awareness of their interesting properties has spurred huge interest: evanescent fields were recently found to have a transverse spin \cite{Bliokh2012,Kim2012,Bliokh2014,Bliokh2015c}, and to exhibit spin-momentum locking \cite{Bliokh2015,Aiello2015,Marrucci2015,Mechelen2016}, leading to a myriad of practical applications in light nano-routing, quantum optics, nonreciprocal devices, optical forces and polarimetry \cite{Rodriguez-Fortuno2013,OConnor2014,Petersen2014,Rodriguez-Fortuno2015,Coles2016,Scheucher2016,Lodahl2017,Espinosa-Soria2017}.~Beyond electromagnetism, evanescent waves have now been found to exhibit analogous properties in other wave fields, such as acoustics \cite{Long2018,Shi2018,Bliokh2019}. This work explores the existence of evanescent waves in the framework of linearised gravity. Inspired by the analogy~to~other~wave fields, we discuss their remarkable properties, which include the transverse spinning of free-falling test masses. Evanescent gravitational waves also imply the excitation of vector and longitudinal components of the wave, which is noteworthy, as the presence of these components in a vacuum is often assumed to signify a deviation from general relativity \cite{Berti:2015itd}. We show that evanescent gravitational waves are not a rare occurrence. They are present and even dominant near any sub-wavelength source of gravitational waves, such as~compact~binary~systems.}
\section{Evanescent waves}
    These can be described using a wave-function that is an eigenmode of the momentum and energy operators. Therefore, they are mathematically identical to plane waves,
$
    \psi(t,\vec{x})=\Psi\exp(\ui \vec{k}\cdot\vec{x}-\ui\omega{t}),
$
    where $\vec{x}$ is the position vector, $t$ is the coordinate time, $\Psi$ is the complex amplitude of this field, $\vec{k}$ is the wave-vector and $\omega$ is the angular frequency. The only difference from travelling plane waves is that the wave-vector, or momentum, will be complex, $\vec{k}=\vec{k}'+\ui\vec{k}''$, with an imaginary component in the direction of the exponential decay. In the case of a vector field, such as the electromagnetic field, the mathematical form of evanescent waves is exactly as above, with $\Psi$ substituted by the electric field amplitude $\vec{E}$. Maxwell equations of electromagnetism impose two conditions on its wave solutions \cite{Jackson1998,Bliokh2015}. Firstly, as every solution to the homogeneous Helmholtz wave-equation, the wave has to be \textit{null-like}, in other words it satisfies the dispersion relation $k_0^2=\vec{k}\cdot\vec{k}=k_x^2+k_y^2+k_z^2$. It is important to stress that for complex-valued wave vector, the quantity $k_0^2=\vec{k}\cdot\vec{k}=|\vec{k}'|^2-|\vec{k}''|^2+2\ui\vec{k}'\cdot\vec{k}''$ is \emph{not} equal to the magnitude of the wave-vector $|\vec{k}|^2=\vec{k}\cdot\vec{k^*}=|\vec{k}'|^2+|\vec{k}''|^2$ \cite{Bliokh2015,Jackson1998,Mechelen2016}. The dispersion relation shows that a wave may surprisingly have $|\vec{k}'|>k_0$, as long as $|\vec{k}''|\neq 0$, demonstrating the mathematical existence of evanescent waves as valid solutions. In a vacuum, this condition can only be satisfied if $\vec{k}'\cdot\vec{k}''=0$, so the direction of decay is necessarily transverse to the direction of propagation. Secondly, the electric field must fulfil the transversality condition $\vec{k}\cdot\vec{E}= 0$ \cite{Bliokh2015,Jackson1998}. This condition restricts the allowed polarisation modes of the wave. It reduces, by one, the three degrees of freedom of vector $\vec{E}$, allowing us to express it as a linear combination of two polarisation basis vectors $\vec{E}=E_1\vu{e}_1+E_2\vu{e}_2$. For plane waves, the condition restricts the electric field to lie on a plane transverse to $\vec{k}$, e.g. two orthogonal linearly polarised waves, or two opposite handedness of circularly polarised waves. For evanescent waves, the same mathematical formulation for the basis vectors can be used, but they become complex-valued \cite{Bliokh2015,Mechelen2016} and, while still fulfilling the condition $\vec{k}\cdot\vec{E}= 0$, the modes acquire longitudinal components of the field ultimately resulting in the remarkable polarisation properties of evanescent waves.
\begin{figure*}[!t]
    \begin{minipage}[t]{0.24\textwidth}
        \includegraphics[width=\textwidth]{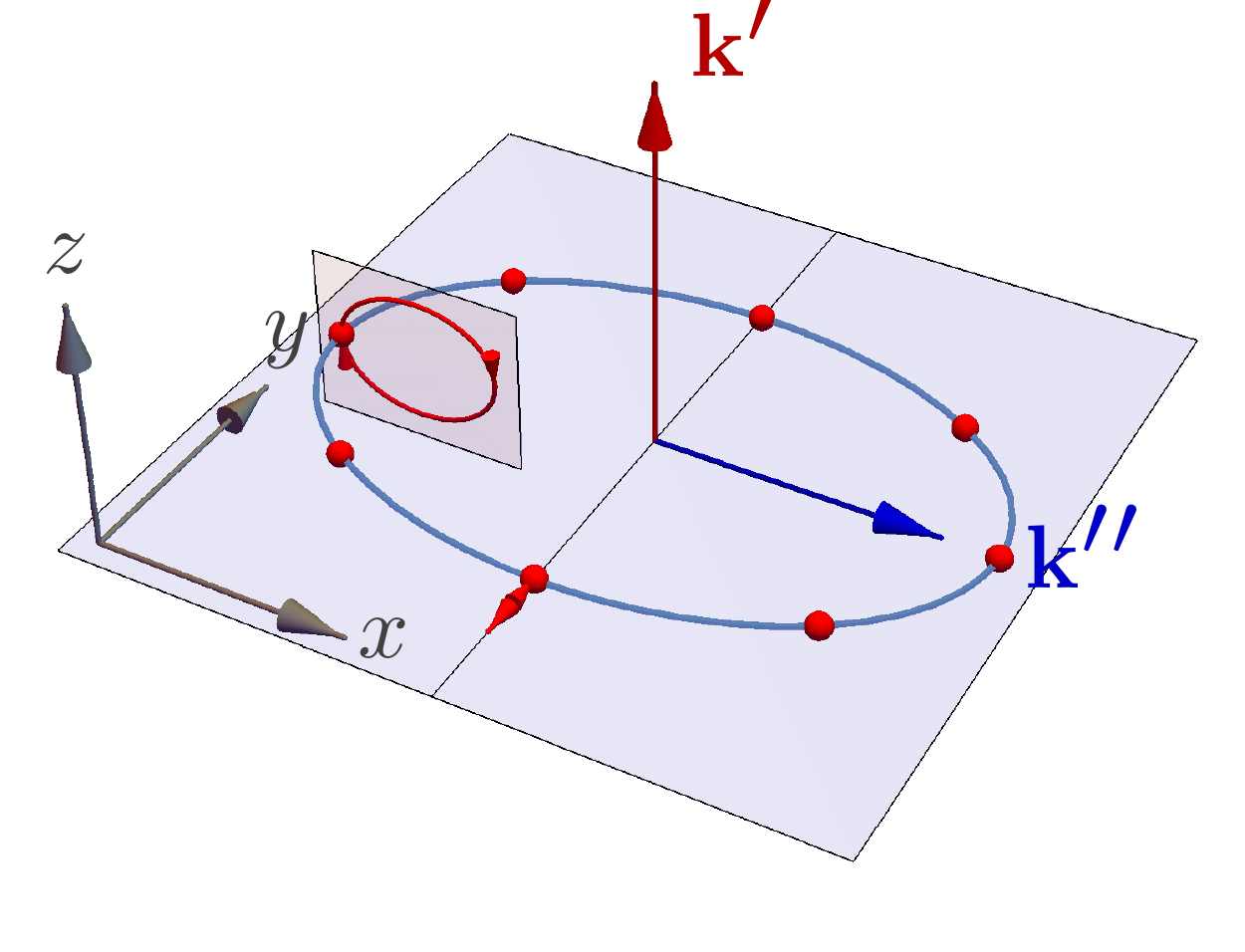}
        {$\phi=0$}
    \end{minipage}
    \begin{minipage}[t]{0.24\textwidth}
        \includegraphics[width=\textwidth]{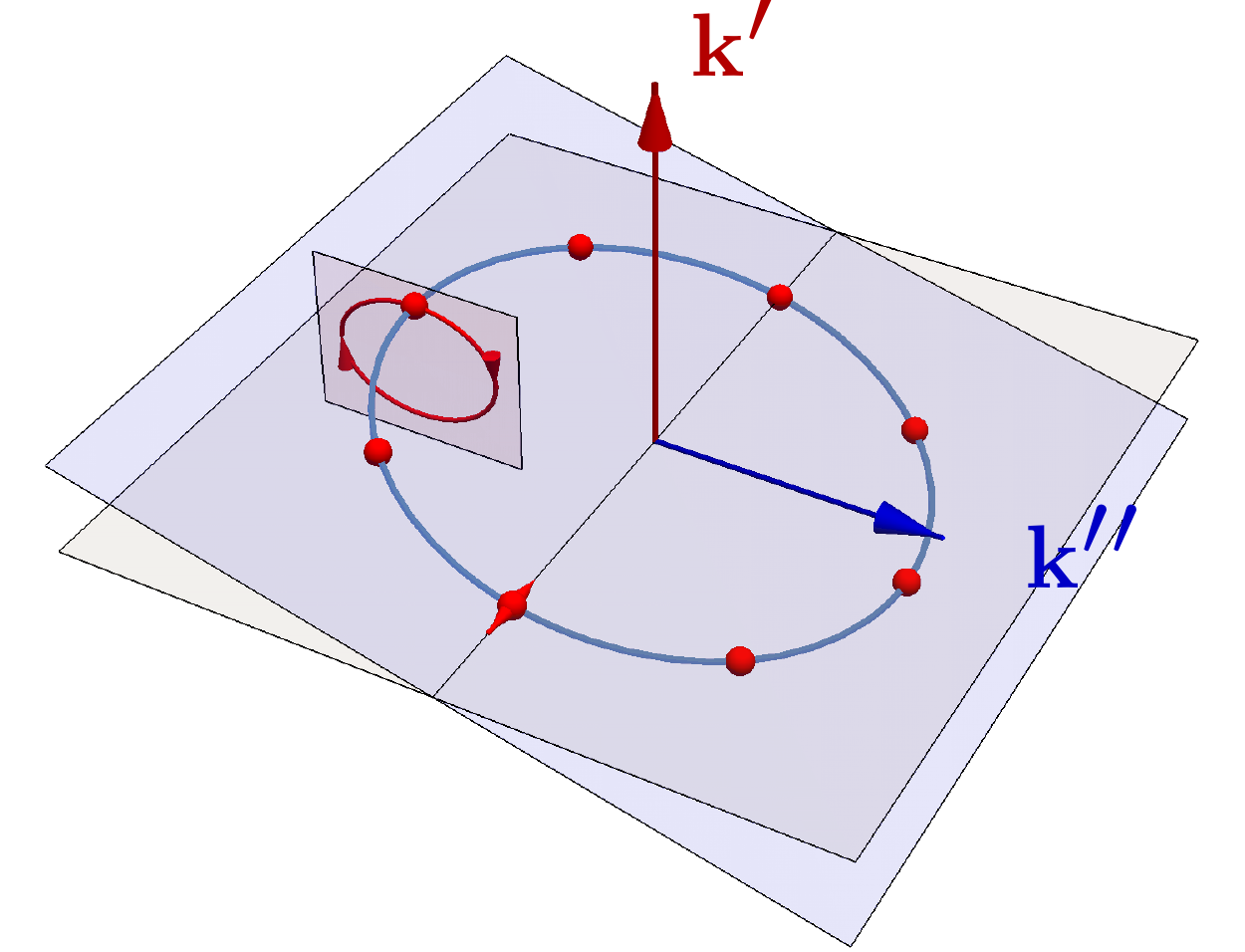}
        {$\phi=\pi/2$}
    \end{minipage}
        \begin{minipage}[t]{0.24\textwidth}
        \includegraphics[width=\textwidth]{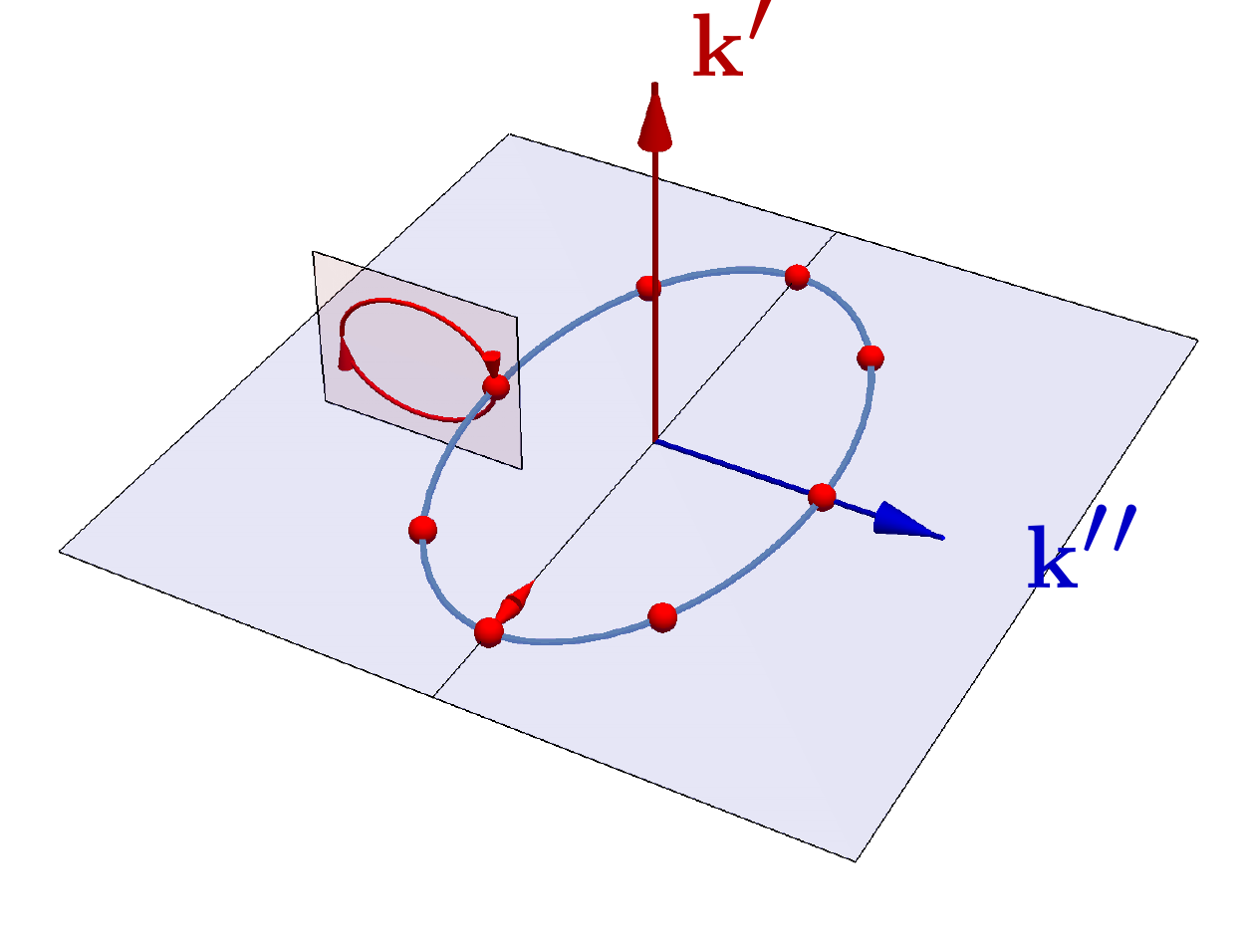}
        {$\phi=\pi$}
    \end{minipage}
        \begin{minipage}[t]{0.24\textwidth}
        \includegraphics[width=\textwidth]{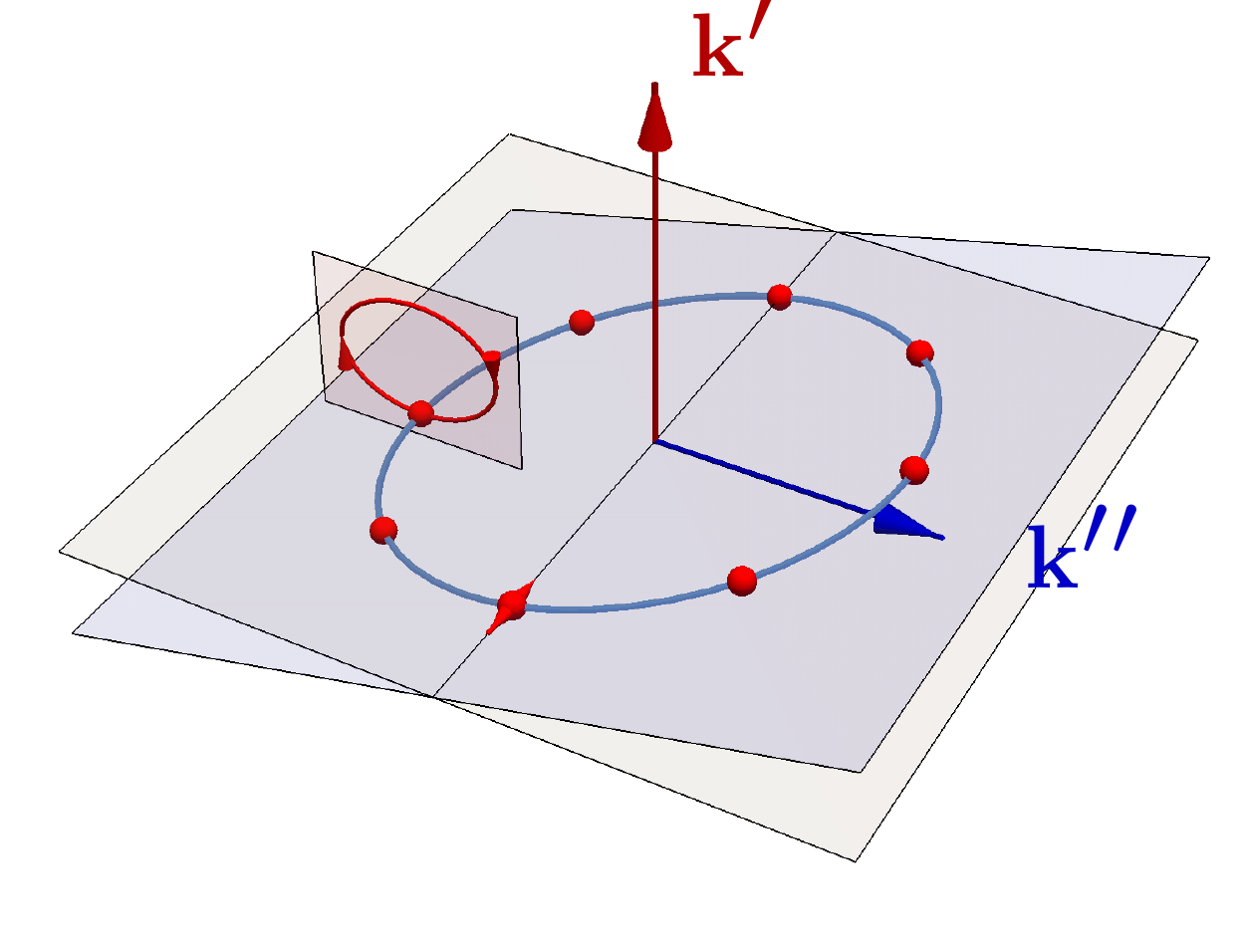}
        {$\phi=3\pi/2$}
    \end{minipage}
    \begin{minipage}[t]{0.24\textwidth}
        \includegraphics[width=\textwidth]{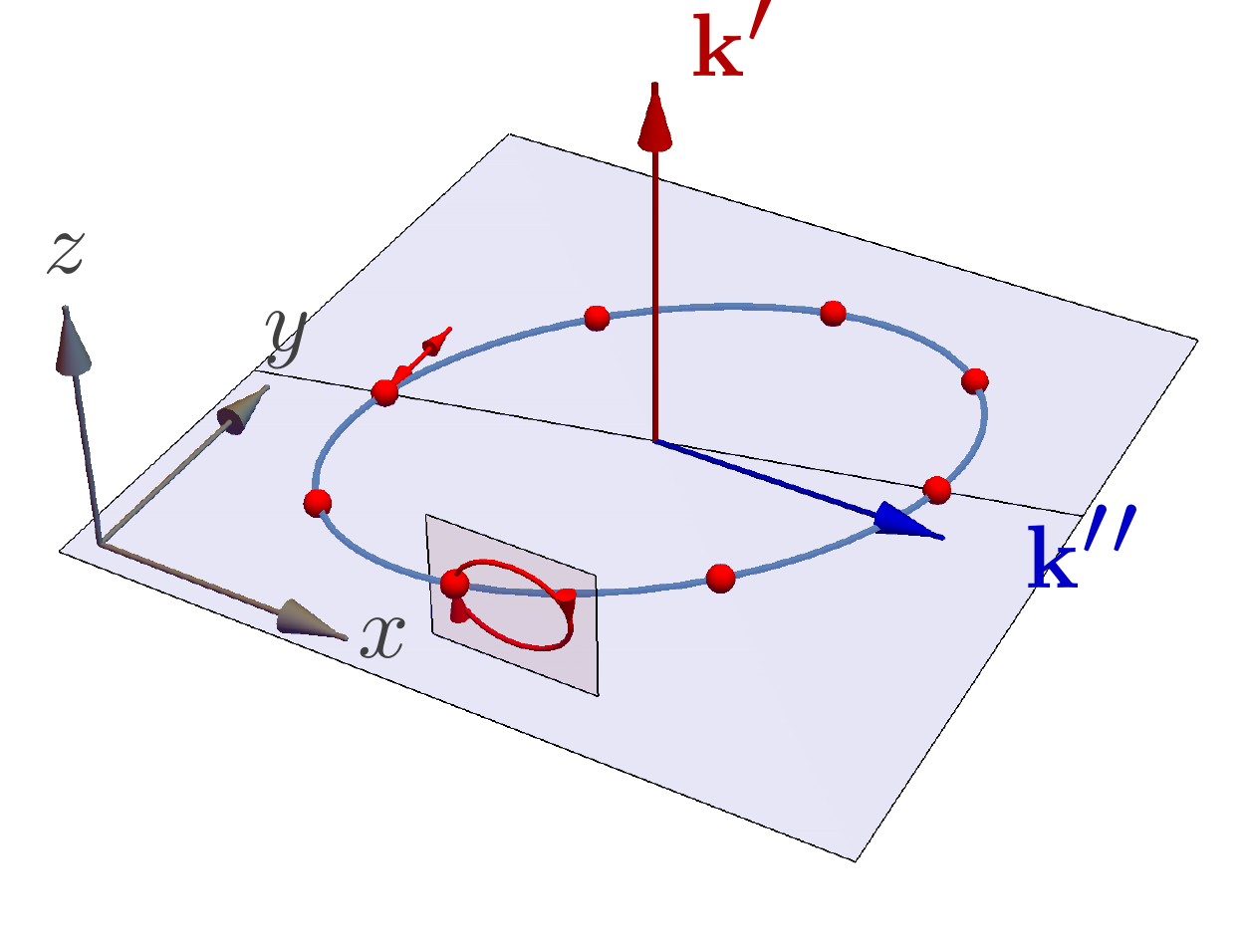}
    \end{minipage}
    \begin{minipage}[t]{0.24\textwidth}
        \includegraphics[width=\textwidth]{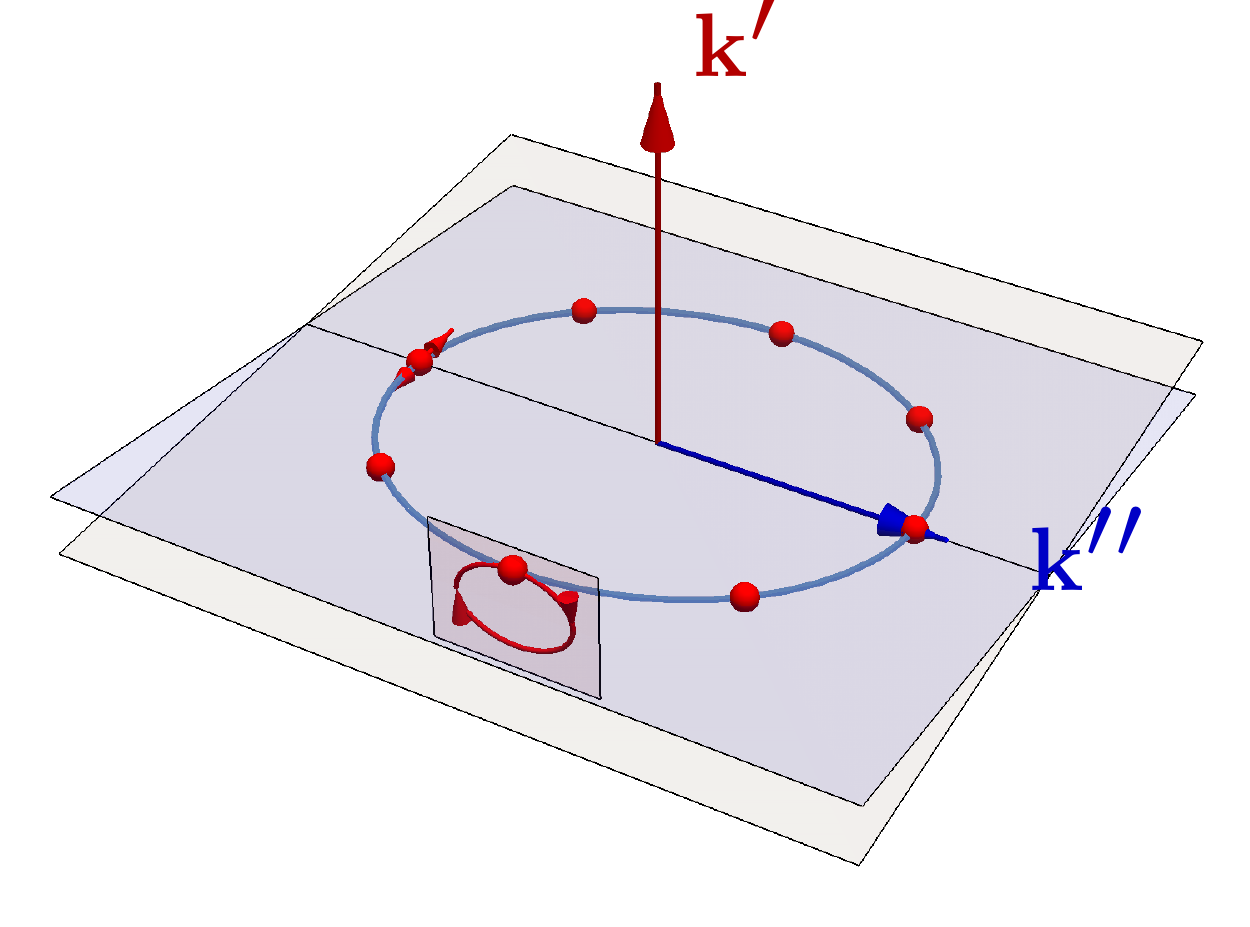}
    \end{minipage}
        \begin{minipage}[t]{0.24\textwidth}
        \includegraphics[width=\textwidth]{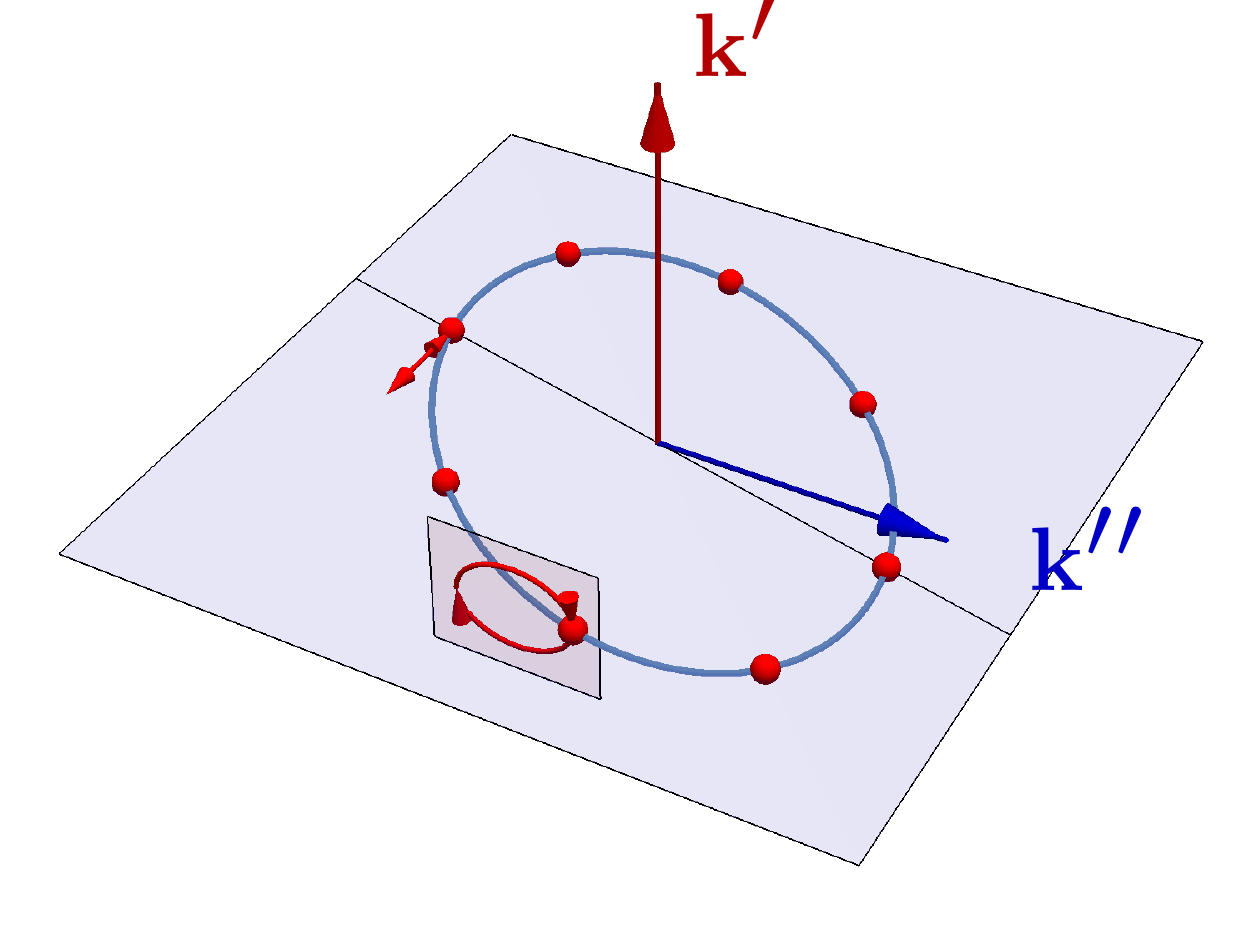}
    \end{minipage}
        \begin{minipage}[t]{0.24\textwidth}
        \includegraphics[width=\textwidth]{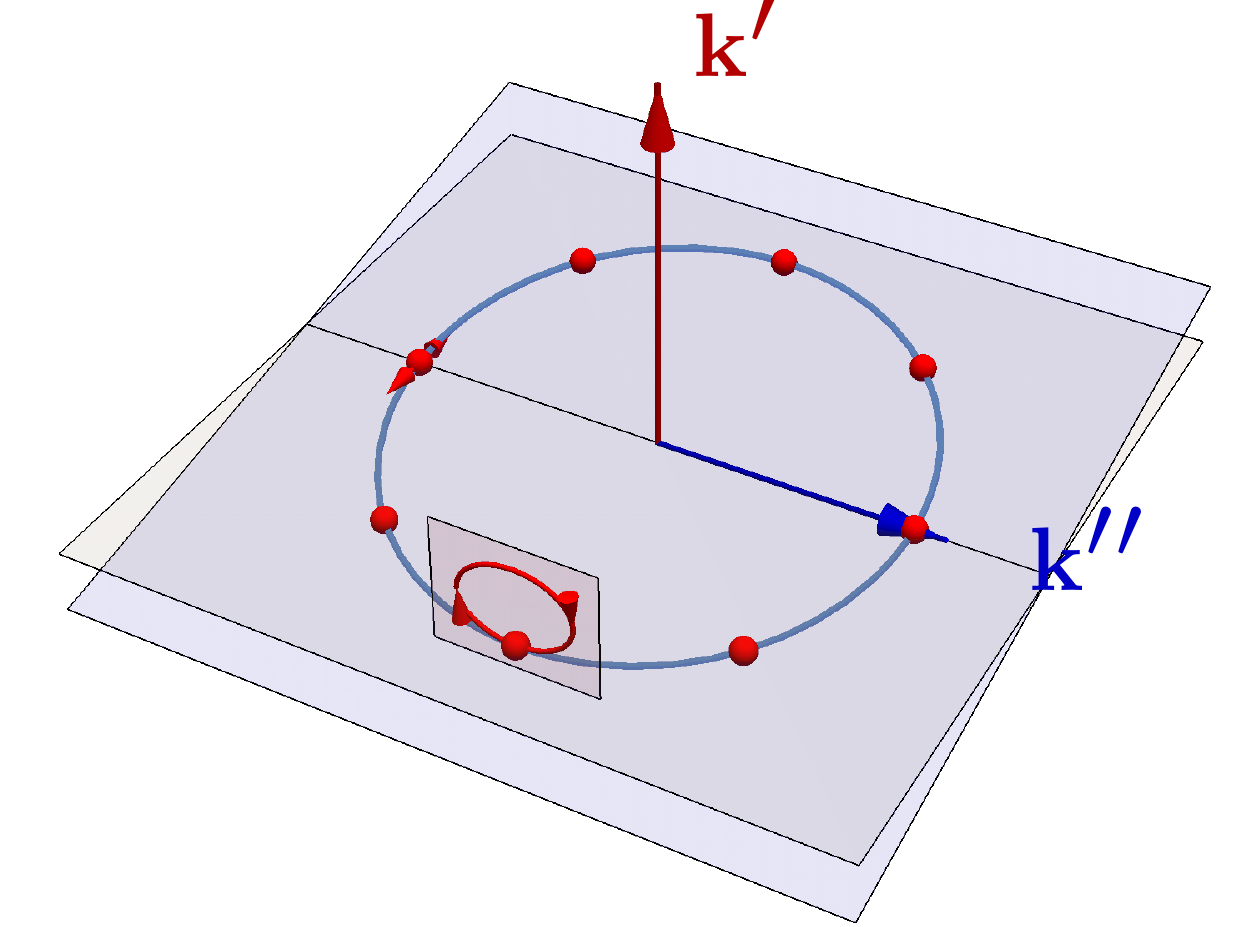}
    \end{minipage}
    \caption{Plot showing the behaviour of a ring of free-falling test masses (red points) at four phases (separated by $\pi/2$) when an evanescent wave of wave vector $\vec{k}=\vec{k'}+\ui\vec{k}''$  passes through the centre in a direction $\vec{k}'$ (red arrow). The wave is decaying in the direction $\vec{k}''$ (blue arrow). To study the local behaviour of the wave, we assume the ring of test masses has a size smaller than the wavelength, such that $\vec{k}\cdot\vec{x}_0\approx 0$. The test masses show not only the usual alternating stretching and squeezing along orthogonal directions characteristic of plane waves, but also individual test masses move in the direction of $\vec{k}'$, oscillating elliptically along a plane parallel to $\vec{k}'$, as shown by the red ellipse. The net effect is that the test masses are always contained on a plane that pivots along an axis perpendicular or parallel to $\vec{k}''$, corresponding to the $+$ (first row) and $\times$ (second row) polarisations respectively. Visit \url{https://youtu.be/DB7mHGqsrLk} for an animated version of this figure.}
    \label{fig:plus}
\end{figure*}
\section{Gravitational waves}
    The theory of linearised gravity describes gravitational plane waves in terms of the metric perturbation symmetric second rank tensor,
\begin{equation}
    {h}_{\mu\nu}(t,\vec{x})={H}_{\mu\nu}\exp(\ui \vec{k}\cdot\vec{x}-\ui\omega{t}).\label{eq:tensor}
\end{equation}   
    The massless wave equation requires the null condition  
\begin{equation}
    k_\mu k^\mu=-k_0^2+k_{x}^2+k_{y}^2+k_{z}^2=0\label{eq:null}
\end{equation} 
    which imposes the dispersion relation. Meanwhile, fixing the gauge to be transverse-traceless implies that, in a vacuum, only the spatial components $h_{ij}$ of ${h}_{\mu\nu}$ are non-vanishing for radiation. The transversality condition is
\begin{equation}
    k^i{h}_{ij}=0,\label{eq:trans}
\end{equation} in this gauge.
    That, together with the trace-less condition ${h}^i{}_i=0$ gives a set of four equations which reduce the original six degrees of freedom of the symmetric matrix ${h}_{ij}$ down to two, therefore restricting it to two allowed polarisation modes. For propagating gravitational plane waves, the two modes may be chosen as the well-known ``plus'' ($+$) and ``cross'' ($\times$) modes, but, in analogy to electromagnetism, these two modes can be extended to the case of complex $\vec{k}$ with the use of a complex basis, as follows.
    Consider an energy and momentum eigenmode gravitational wave, \cref{eq:tensor}, travelling in the $z$-direction and decaying in $x$ (without loss of generality due to the fact that $\vec{k}'\cdot\vec{k}''=0$). This implies a complex wave-vector $\vec{k}=k_0\mqty(\ui\alpha&,0&,\kappa)$ where $\alpha$ and $\kappa$ are both real. The null condition requires that $1=\kappa^2-\alpha^2$. Any propagating mode can then be expressed as a linear combination of two {\it complex} polarisation modes (see \cref{sec:localdes}):
\begingroup % keep the change local
\setlength\arraycolsep{2.5pt}
\begin{equation}\label{eq:evan}
     {H}_{ij}= h_+\Pmqty{\kappa^2&0&-\ui\alpha\kappa\\0&-1&0\\-\ui \alpha\kappa&0&-\alpha^2}+h_\times\Pmqty{0& \kappa&0\\  \kappa&0&-\ui\alpha\\0&-\ui\alpha&0}.
\end{equation}
\endgroup
\begin{table}[!b]
    \centering
    \caption{Real polarisation basis decomposition}
    \begingroup % keep the change local
\renewcommand*{\arraystretch}{1.6}
\setlength{\tabcolsep}{5pt}
    \begin{tabular}{ | c | c | l | }
    \hline
    $\mathcal{H}_+$ & $h_+\qty(1+{\alpha^2}/{2})$& Plus mode \\
    \hline
    $\mathcal{H}_\times$ & $h_\times\sqrt{1+\alpha^2}$ & Cross mode\\
    \hline
    $\mathcal{H}_1$ & $-h_+\ui\alpha\kappa$& Vector-$x$ mode\\
    \hline
    $\mathcal{H}_2$ & $-h_\times\ui\alpha$ & Vector-$y$ mode\\
    \hline
    $\mathcal{H}_3$ & $-h_+\alpha^2$ & Longitudinal mode\\
    \hline
    $\mathcal{H}_0$ & $h_+{\alpha^2}/{2}$& Breathing mode \\
    \hline
    \end{tabular}
    \endgroup
    \label{tab:decompos}
\end{table}
    The complex nature of these amplitudes accounts for the amplitude and phase of each component. Note that these two modes reduce to the usual gravitational transverse ``plus'' and ``cross'' modes when $\kappa\rightarrow1$, and correspondingly $\alpha\rightarrow0$ due to the null condition. Crucially, in this basis, components which are not transverse to the direction of propagation given by $\vec{k}'$ are present, even though the transversality condition $k^i{h}_{ij}=0$ is satisfied. This is analogous to the appearance of longitudinal fields in evanescent electromagnetic waves. To show this clearly, we may decompose our basis in terms of the real polarisation basis ${H}_{ij}=\sum\mathcal{H}_A{e^A_{ij}}$ (see \cref{sec:A}), summed over $A\in\qty{+,\times,0,1,2,3}$ as shown in \cref{tab:decompos}. Thus, evanescent waves in a vacuum can excite the vector and scalar modes even in general relativity---complicating efforts to use the detection of such modes as a smoking gun evidence for modified gravity theories \cite{Berti:2015itd}. Nevertheless, we emphasise that these are \emph{not} additional modes as there are only two effective propagating degrees of freedom---the key point is that the components of these real modes are correlated. 
\section{Motion of test masses and transverse spin}
    To study the effects of the wave, we can consider a cloud of freely falling test masses surrounding a fixed point. If the cloud is small compared to the wavelength, the effect of the wave on one of the particles of the cloud can be examined using the geodesic deviation equation.
$
    \ddot{x}^i=-{R}^i{}_{0j0}(t){x}^j=\frac{1}{2}\Re[\ddot{h}^i{}_j(t)]{x}^j,
$
    where  ${x}^i(t)$ are coordinates representing instantaneous proper positions of the free-falling mass with respect to a fixed point, ${R}^\mu{}_{\nu\rho\sigma}(t)$ is the Riemann curvature tensor evaluated at the fixed point, and the dot represents partial derivative with respect to time \cite{Misner1973}. Given initial positions ${x}_0^i$, this equation has a unique solution
\begin{equation}\label{eq:traject}
    {x}^i(t)={x}_0^i+\delta{x}^i(t)={x}_0^i+\frac{1}{2}\Re[{h}^i{}_j(t)]{x}_0^j.
\end{equation}
    The displacement, for eigenmode \cref{eq:tensor}, can be written 
\begin{equation*}
    \delta{x}^i(t)=\frac{1}{2}\qty[\Re\qty({H}^i{}_j)\cos(\omega{t})
    +\Im\qty({H}^i{}_j)\sin(\omega{t})]{x}_0^j,
\end{equation*}
    which is the parametric equation of an ellipse. Therefore, each test mass will move along a fixed ellipse with centre at ${x}_0^i$ and semi-axes defined by two \textit{conjugate diameter vectors} $\frac{1}{2}\Re\qty({H}^i{}_j){x}_0^j$ and $\frac{1}{2}\Im\qty({H}^i{}_j){x}_0^j$. If we consider propagating non-evanescent plane waves with $\kappa = 1$ in \cref{eq:evan}, the elliptical orbit becomes a line segment perpendicular to the propagation vector $\vec{k}$. Thus, under the influence of non-evanescent gravitational waves, test masses oscillate within a fixed plane perpendicular to the direction of the wave vector $\vec{k}$, alternately stretching and squeezing along perpendicular directions in the distinct $+$ and $\times$ pattern. Under the influence of evanescent waves, test masses show the same pattern but acquire an additional movement in the longitudinal direction $\vec{k}'$. When both movements are combined coherently, the masses follow elliptical trajectories on planes parallel to $\vec{k}'$. For high values of $\kappa \rightarrow\infty$, corresponding to more confined evanescent waves, some of the trajectories become perfect circles. The net effect of this motion is that the test masses carry out the usual $+$ and $\times$ oscillations, but they do so on a plane that is not perpendicular to $\vec{k}'$ at all times, and instead pivots---like the rocking motion of a playground see-saw (\cref{fig:plus} or \url{https://youtu.be/DB7mHGqsrLk} for an animated version). Note that the imaginary component of the wave-vector breaks the rotational symmetry of the two~modes. The elliptical movement of the masses is hugely reminiscent of the transverse spin of evanescent electromagnetic and acoustic waves. 
    As described in \cref{sec:tspin}, one may calculate the spin angular momentum density of a gravitational wave as
\begin{equation}
    \vec{S}=\frac{W}{\omega}\qty[2\sigma\frac{\vec{k}'}{\qty|\vec{k}'|}+2\frac{\vec{k}'\cp\vec{k}''}{\qty|\vec{k}'|^2}],\label{eq:tspin}
\end{equation} 
    where $\sigma$ is the normalised third Stokes parameter or \textit{helicity parameter}, equal to $\pm1$ for purely circularly polarized waves, defined as 
$\sigma=2\Im{h_+^\ast h_\times}/(\qty|h_+|^2+\qty|h_\times|^2)$ and $W$ is the time-averaged energy density \cite{Barnett2014}
\begin{equation}
    W=\dfrac{c^2}{128\pi G}\qty[\partial_t h^\ast_{ij}\partial_t h^{ij}+c^2
    \epsilon_{jmn} \epsilon^{jkl}
    \partial_l h^\ast_{ik} 
    \partial^m h^{in}
    ].
\end{equation}%
    From \cref{eq:tspin}, the expected spin-2 nature of the longitudinal intrinsic angular momentum of gravitational waves appears in the first term -- in clear analogy to the spin-1 nature of the electromagnetic~wave~\cite{Bliokh2015c}. However, the novel second term represents an intrinsic \emph{transverse spin} with a value of $+2\alpha/\kappa$. (Interestingly, this is identical to the transverse spin of acoustic waves \cite{Bliokh2019}, and twice that of electromagnetic waves). Spin-momentum locking is manifest because reversing the direction of $\vec{k}'$ also changes the sign of $\vec{S}$. 
\section{Generation of evanescent fields}
    Having described evanescent gravitational waves as a valid solution to the vacuum wave equation, we now discuss their occurrence in nature. A straightforward way to produce evanescent waves is to use the phenomenon of total internal reflection. Consider an electromagnetic plane wave incident on an interface between two media with a different index of refraction. If the angle of incidence is greater than the critical angle then, using conservation of $\vec{k}$ parallel to the interface, i.e. Snell's law, the wave vector of the refracted wave will be complex \cite{Fornel2001,Jackson1998}. For gravitational waves, even though theoretically conceivable, this possibility seems to be physically unrealistic since scattering by matter \cite{Thorne1983} is negligible \footnote{We note that for gravitational waves in matter, the angle between $\mathbf{k}'$ and $\mathbf{k}''$ will be determined by the dispersion relationship which will no longer be the null condition.}. As a consequence, the medium which could refract or reflect gravitational waves requires either exotic material with large shear modulus or shear viscosity~\cite{Press1979} or an array of tightly packed sufficiently compact objects (like black holes or neutron stars)~\cite{Thorne1983}. However, we need not consider such exotic scenarios to observe evanescent waves. Evanescent gravitational fields, in analogy to any other type of wave, must be present in the near-field zone of any sub-wavelength source \cite{Fornel2001}. This fact can be understood via the position-momentum Fourier properties of any wave $\Delta{x}\Delta{k}\geq1/2$. A localised sub-wavelength source with $\Delta x < \frac{1}{2 k_0}=\frac{\lambda}{4 \pi}$ necessarily has a wide range of momentum values, i.e. range of wave-vectors, which extend beyond the wave-number of free space $\Delta k_x > k_0$. As discussed earlier $|\mathbf{k}'|>k_0$ implies, from the dispersion relation in Eq.\ (\ref{eq:null}), that  $|\mathbf{k}''|> 0$, constituting evanescent components. The decay of these components when far from the source is responsible for the diffraction limit in far-field imaging. 
\section{Momentum space representation}
    To show that there are evanescent components near a sub-wavelength source of gravitational radiation, we will use the angular spectrum representation. This is a standard tool for studying wave-fields in homogeneous media and is widely used in nanophotonics to study scattering, beam propagation, focusing, holography, and many other phenomena \cite{Mandel1995}. The main idea is that, in general, solutions to the wave equation are not momentum eigenmodes with a well-defined wave-vector $\vec{k}$ as in \cref{eq:tensor}, but rather a distribution of them constituting a continuous spectrum. The generalisation of this representation to rank-2 tensor gravitational waves can be made, as shown below. Suppose we know a field ${h}_{ij}(\vec{r},t)$ at any point. We may assume it is time-harmonic without loss of generality as we can always perform a temporal Fourier transform. We can consider this field in a plane $z=\text{constant}$ which is transverse to an arbitrary $z$-direction. In this plane, we can write the field as a 2D inverse spatial Fourier transform \cite{Mandel1995}:
\begin{eqnarray*}
    {h}_{ij}\qty(\vec{r},t)=&\displaystyle\iint\limits_{-\infty}^{\phantom{--}\infty}{{\tilde{h}}_{ij}(k_x,k_y)}\ue^{\ui\qty(k_xx+k_yy+k_z\abs{z}-\omega t)}\dd{k_x}\dd{k_y}.
\end{eqnarray*}
 Since the wave satisfies the wave equation (and hence the null condition \cref{eq:null}) and we consider these waves to be time-harmonic, we can always uniquely (up to a sign) find $k_z$ for each pair of $k_x$ and $k_y$. By considering the tensor ${\tilde{h}}_{ij}(k_x,k_y)$ as a momentum eigenmode and we express it as a superposition of only two tensor modes as in \cref{eq:evan}:
\begin{equation}
    \tilde{h}_{ij}(k_x,k_y)=h_+(k_x,k_y){e}^+_{ij}(\vec{k})+ h_{\times}(k_x,k_y){e}^{\times}_{ij}(\vec{k}),\label{eq:decompost}
\end{equation}
    where ${e}^{+}_{ij}(\vec{k})$ and ${e}^{\times}_{ij}(\vec{k})$ are the complex basis tensors introduced in \cref{eq:evan}, generalised for arbitrary direction of vector $\vec{k}$ (see \cref{sec:A}). Therefore, a pair of scalar complex-valued angular spectra $h_+(k_x,k_y)$ and $h_\times(k_x,k_y)$ completely describe the source in momentum (and also real) space. In the region $k_x^2+k_y^2\leq k_0^2$ these two spectra correspond to real $k_z$ (plane waves propagating to the far-field). The region $k_x^2+k_y^2>k_0^2$, corresponds to an imaginary $k_z$ associated with evanescent near field components whose amplitude decays exponentially as $\abs{z}$ is increased.
    The space around a source where evanescent components dominate is known in electromagnetism as the \textit{reactive near-field} zone (approximately $r \lesssim \lambda/2\pi$ \cite{Balanis2016}), and will also exist for gravitational waves. The linearised gravity approximation can still apply in this region, as its validity breaks much closer to the source ($r\sim 5R_\mathrm{S}$ \cite{Thorne1983}, where $R_\mathrm{S}$ is Schwarzschild radius).

\begin{figure}[!t] 
    \centering
    \includegraphics[width=\linewidth]{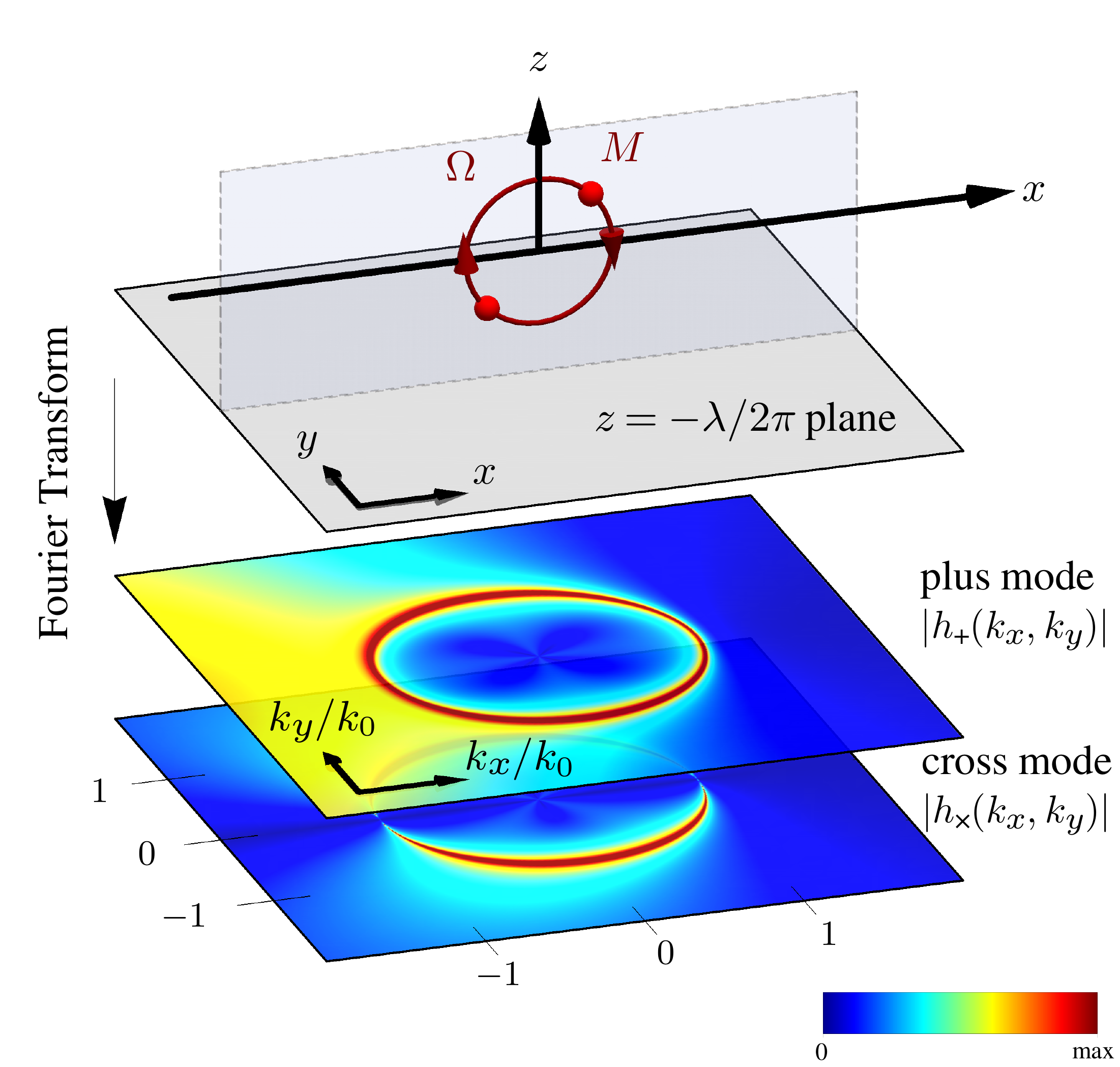}
    \caption%
    {Momentum space (angular spectrum) representation of the quadrupole radiation as felt on the plane $z=-\lambda/2\pi$. The top (bottom) colour-map represents the complex amplitude of the plus (cross) polarisation mode in momentum space. Notice that the spectra show a near-field directionality similar to a circularly polarised electric dipole described in \citet{Picardi2017}, a signature of spin-momentum locking.}\label{fig:xz}
\end{figure}

% \section{Momentum space representation of a quadrupole}
    As the simplest example, in electromagnetism, we can find evanescent fields near a radiating electric dipole \cite{Novotny1997,Fornel2001,Picardi2018}. Due to the quadrupolar nature of gravitational waves, we expect there will be evanescent fields near a radiating gravitational quadrupole. We consider a binary system of compact objects with same mass $M$ separated by a distance $d$ in a stable circular orbit with frequency $\Omega$ around a common centre of mass (\cref{fig:xz}).
    Furthermore, we assume that the speed of the masses is not relativistic $v\ll c$ (note that this directly implies that this is a sub-wavelength source $\vec{k}\cdot\vec{d}\ll 1 \Leftrightarrow d\ll \lambda/2\pi$). For such source, there is always a region ($ d \ll r \lesssim \lambda/2\pi$) which is within the near-field region $r \lesssim \lambda/2\pi$, but is far enough from the two masses that they can be taken together as a localised point source $r\gg d$. In this region, the well-known quadrupole solution to the linearised Einstein equation applies. This solution, in frequency space, is given by:
\begin{equation*}
    \bar{h}_{ij}(\omega, r) = \frac{G}{4c^4}\frac{\omega^2}{r}e^{i\omega r/c}q_{ij}\label{eq:hijspherical},
\end{equation*}
    where the frequency $\omega=k_0c=2\Omega$ and
\begin{equation*}
    {q}_{ij}= Md^2\frac{1}{2}\Pmqty{-1&0&-\ui\\0&0&0\\-\ui&0&1}
\end{equation*}
    is a constant tensor. This form allows us to reproduce identical mathematical steps as taken in \citet{Picardi2017} for an electromagnetic dipole to find the angular spectra of the gravitational quadrupole, details of this calculation are shown in the \cref{sec:genesis}. As a result, we present expressions for the two scalar amplitudes which represent separately the two polarisation mode angular spectra of the gravitational quadrupole;
\begin{subequations}
\begin{eqnarray}
    h_+(k_x,k_y)&=\displaystyle\frac{\ui G}{16\pi c^2}\frac{k_0^2}{k_z}{q}_{ij}{{e}}_+^{ij}(k_x,k_y),\\
    h_\times(k_x,k_y)&=\displaystyle\frac{\ui G}{16\pi c^2}\frac{k_0^2}{k_z}{q}_{ij}{{e}}_\times^{ij}(k_x,k_y).
\end{eqnarray}
\end{subequations}
    These amplitudes contain all the information necessary to reconstruct the fields of the quadrupole source at every location in space, including its near field. The complex amplitudes of the two spectra $|h_+|$ and $|h_\times|$ are plotted in \cref{fig:xz}
    after propagating them to a plane $z = -\lambda/2\pi$ via the transfer function $\ue^{- \ui k_z z}$. We see that for the considered sub-wavelength distance, there is a strong presence of waves with $\vec{k}$ in the region $k_x^2+k_y^2>k_0^2$, corresponding to evanescent waves. A similar procedure can be repeated for any other source of gravitational waves, to show that in the near field of any source there will be a full spectrum of evanescent waves.
\section{Conclusions}
    In recent years, evanescent waves and their properties have raised considerable interest in optics and acoustics. This letter is the first work dedicated to the study of these in the context of gravity. This required extending the formalism from vector to tensor modes. We have found that not only evanescent gravitational waves can exist, but also that they are not exotic phenomena and one can expect them in the near zone of any source of gravitational waves. In analogy with electromagnetic and acoustic waves, gravitational waves also possess non-trivial polarisations associated with a transverse spin and spin-momentum locking. Another implication of the existence of evanescent gravitational waves is that, even if non-tensorial modes of gravitational waves in a vacuum are detected, this does not necessarily contradict general relativity as they may originate from an evanescent field: one can check whether the polarisation modes are correlated. Non-tensorial modes may also originate from the coherent superposition of two propagating plane waves arriving simultaneously at a detector, whose combined polarisation can be locally identical to that of an evanescent wave (see \cref{sec:localdes}), in analogy to the transverse spin that appears in electromagnetic two-wave interference \cite{Bliokh2015c}. Direct detection of near field evanescent gravitational waves is unlikely -- e.g. the near zone for LIGO and LISA type detectors would be $10^{-7}$ and $10^{-2}$ A.U. respectively. In the meantime, due to the effect of evanescent gravitational waves on test particles, there are some potential options for indirect detection. For example, charged particles moving on an elliptical trajectory due to the evanescent fields near the source of gravitational waves should radiate electromagnetic radiation whose polarisation signature we could detect.
%TC:ignore 
\begin{acknowledgments}
FJRF is supported by European Research Council Starting Grant ERC-2016-STG-714151-PSINFONI. EAL is supported by STFC AGP-AT Grant ST/P000606/1. SG is supported by Scholarship of the city of Ostrava.
\end{acknowledgments}
%%%%%%%%%%%%%%%%%%%%%%%%%%%%%%%%%%%%%%%%%%%%%%%%%%%%%%%%%%%%%%%%%%%%%%%%%%%%%%%%%%%%%%%%%%%%%%%%%%%%%%%%%%%%%%%%%%%%%%%%%%%
\appendix
%%%%%%%%%%%%%%%%%%%%%%%%%%%%%%%%%%%%%%%%%%%%%%%%%%%%%%%%%%%%%%%%%%%%%%%%%%%%%%%%%%%%%%%%%%%%%%%%%%%%%%%%%%%%%%%%%%%%%%%%%%%
\section{Polarisation basis of evanescent fields}\label{sec:A}
    In linearised gravity, gravitational waves are often expressed in the transverse-traceless (TT) gauge in terms of the traceless symmetric metric perturbation 
    \begin{equation}
    h_{\mu\nu}(t,{\vec{x}}) = H_{\mu\nu}\exp(\ui{\vec{k}}\cdot {\vec{x}}-\ui\omega t)
    \end{equation}
    where ${\vec{k}}$ is a three vector pointing in the spatial dimensions and $\omega$ is the frequency, and the Greek indices run from $0$ to $3$. In the TT gauge, there always exist an appropriate gauge transformation such that  $h_{0\mu}=0$ (even in the presence of evanescent waves), hence in vacuum one can represent the perturbation in its spatial components $h_{ij}$ ($i,j=1,2,3$) in the Cartesian coordinate basis.
    
    In vacuum, it is well known that gravitational wave solutions of the Einstein equation possess two polarisation modes, 
    \begin{equation}
    H_{ij} = h_+e^{+}_{ij}({\vec{k}})+ h_{\times}e^{\times}_{ij}({\vec{k}})
    \end{equation}
    where $e^{+}_{ij}({\vec{k}})$ and $e^{\times}_{ij}({\vec{k}})$ are the + and $\times$ polarisation modes, which depend on the wave vector ${\vec{k}}$. Given any ${\vec{k}}$ with components $k_i=(k_x,k_y,k_z)$ and $k_0\equiv \sqrt{{\vec{k}}\cdot {\vec{k}}}$, one can construct these basis modes via the following construction
    \begin{align}
      e^{+}_{ij}({\vec{k}}) &= e^{\phi}_{i}({\vec{k}})e^{\phi}_{j}({\vec{k}})-e^{\theta}_{i}({\vec{k}})e^{\theta}_{j}({\vec{k}}),\notag\\
      e^{\times}_{ij}({\vec{k}}) &= e^{\phi}_{i}({\vec{k}})e^{\theta}_{j}({\vec{k}})+e^{\phi}_{i}({\vec{k}})e^{\theta}_{j}({\vec{k}}). \label{eqn:basis}
    \end{align}
    Unit co-vectors $e^{\phi}_{i}$ and $e^{\theta}_{i}$ are transverse to the wave-vector ($e^{\phi}_{i}({\vec{k}})k^i=e^{\theta}_{i}({\vec{k}})k^i=0$). They can be written in terms of components of the wave vector as
    \begin{align*}
    e^{\theta}_{i}({\vec{k}})&=\frac{1}{\sqrt{k_x^2+k_y^2}}\Pmqty{-k_y,&k_x,&0}^\intercal,\\
    e^{\phi}_{i}({\vec{k}})&=\frac{1}{k_0\sqrt{k_x^2+k_y^2}}\Pmqty{k_x k_z,& k_y k_z,& -k_x^2-k_y^2}^\intercal,
    \end{align*}%
    which, when ${\vec{k}}$ is real, correspond to the usual basis vectors in spherical coordinates.
    For any arbitrary {$\mathbf{k}$} vector, including complex valued ones, we can still use \cref{eqn:basis} to construct the polarisation basis as
    \begin{align}
     e^{+}_{ij}({\vec{k}}) &= \frac{1}{k_0^2}\Pmqty{ 
          \frac{k_x^2k_z^2-k_y^2k_0^2}{k_x^2+k_y^2} & \frac{k_0^2+k_z^2}{k_x^2+k_y^2} {k_x}  k_y & -{{k_x} {k_z}}\\ \frac{k_0^2+k_z^2}{k_x^2+k_y^2} {k_x}  k_y & \frac{k_y^2k_z^2-k_x^2 k_0^2}{k_x^2+k_y^2} & -{ k_y {k_z}}\\ -{{k_x} {k_z}} & -{ k_y {k_z}} & {k_x^2+k_y^2}  
    }, \label{eqn:eplus}\\
     e^{\times}_{ij}({\vec{k}}) &= \frac{1}{k_0}\Pmqty{ 
    \frac{2{k_x}{k_y}}{k_x^2+k_y^2} {k_z} & \frac{k_x^2-k_y^2}{k_x^2+k_y^2} {k_z} & \phantom{-}k_y\\ \frac{k_x^2-k_y^2}{k_x^2+k_y^2} {k_z} & \frac{2{k_x}{k_y}}{k_x^2+k_y^2} {k_z}& -k_x\\ \phantom{-}k_y & -k_x & \phantom{-}0
    }. \label{eqn:ecross}
    \end{align}
    
    When waves are evanescent, the wave-vector $k^i$ is complex in general, and the transversality condition also implies that the basis vectors $e^{\theta}_i({\vec{k}})$ and $e^{\phi}_i({\vec{k}})$ are also complex. Hence, the polarisation modes \cref{eqn:eplus} and \cref{eqn:ecross} are also in general complex. However, it is easy to show that both polarisation modes are still solutions of the gravitational wave equation as long as the null condition $\omega^2/c^2=k_0^2=k_x^2+k_y^2+k_z^2$ is satisfied, and therefore form a valid basis in general. %
    In \cref{tab:decompos} in the main text we decomposed the wave into a real polarisation basis  ${H}{_{ij}}=\sum\mathcal{H}_A {e}{^A_{ij}}({\vec{k}}={\vec{k}}')$, summed over $A\in\{+,\times,0,1,2,3\}$. Note that it is decomposed along ${\vec{k}}'$ which is the real part of the wave vector, $\mathcal{H}_A$ is an amplitude of the corresponding mode $A$ and ${e}{^A_{ij}}$ are basis tensors with only real components see \cref{tab:decomposbasis}. 
    \begin{table}[ht]
        \centering
        \caption{Definition of the real polarisation basis, here $\hat{k}'_{i}$ are the components of a unit vector in the direction of $\vec{k}'$ \cite{Philippoz2017}.}
        \begingroup % keep the change local
    \renewcommand*{\arraystretch}{1.8}
    \setlength{\tabcolsep}{5pt}
        \begin{tabular}{ | l | l | }
        \hline
             $e^{+}_{ij}({\vec{k}'}) = e^{\phi}_{i}({\vec{k}'})e^{\phi}_{j}({\vec{k}'})-e^{\theta}_{i}({\vec{k}'})e^{\theta}_{j}({\vec{k}'})$& Plus mode \\\hline
             $e^{\times}_{ij}({\vec{k}'}) = e^{\phi}_{i}({\vec{k}'})e^{\theta}_{j}({\vec{k}'})+e^{\phi}_{i}({\vec{k}'})e^{\theta}_{j}({\vec{k}'})$ & Cross mode\\\hline
             $e^{1}_{ij}({\vec{k}'}) = e^{\phi}_{i}({\vec{k}'})\hat{k}'_{j}+\hat{k}'_{i}e^{\phi}_{j}({\vec{k}'})$& Vector-$x$ mode\\\hline
             $e^{2}_{ij}({\vec{k}'}) = e^{\theta}_{i}({\vec{k}'})\hat{k}'_{j}+\hat{k}'_{i}e^{\theta}_{j}({\vec{k}'})$ & Vector-$y$ mode\\\hline
             $e^{3}_{ij}({\vec{k}'}) = \hat{k}'_{i}\hat{k}'_{j}$ & Longitudinal mode\\
            \hline
             $e^{0}_{ij}({\vec{k}'}) = e^{\phi}_{i}({\vec{k}'})e^{\phi}_{j}({\vec{k}'})+e^{\theta}_{i}({\vec{k}'})e^{\theta}_{j}({\vec{k}'})$& Breathing mode \\\hline
        \end{tabular}
        \endgroup
        \label{tab:decomposbasis}
    \end{table}
%%%%%%%%%%%%%%%%%%%%%%%%%%%%%%%%%%%%%%%%%%%%%%%%%%%%%%%%%%%%%%%%%%%%%%%%%%%%%%%%%%%%%%%%%%%%%%%%%%%%%%%%%%%%%%%%%%%%%%%%%%%
\section{Motion of test masses under Evanescent Waves}
    In this section, we calculate the motion of test masses in the presence of a single mode of evanescent gravitational wave, and show that the loci of test masses are ellipses (as opposed to straight lines in plane waves). In vacuum ${\vec{k}'}\cdot{\vec{k}''} = 0$, so without loss of generality we may orient our axes to consider a wave vector
    \begin{equation}
      {\vec{k}}=k_0 
      \Pmqty{\kappa_x \\ 0 \\ \kappa_z \\}
      ~,~1=\kappa_x^2+\kappa_z^2,
    \end{equation}
    where the second equation imposes the null-like condition. Using \cref{eqn:eplus} and \cref{eqn:ecross}, the polarisation modes for this wave are then
    \begin{align}
    e_{ij}^{+} &= 
    \Pmqty{\kappa_z^2 & 0 & -\kappa_x\kappa_z \\
    0 & -1 & 0 \\
    -\kappa_x\kappa_z & 0 & \kappa_x^2 },\\
    e_{ij}^{\times} &=\Pmqty{
    0 & \kappa_z & 0 \\
    \kappa_z & 0 & -\kappa_x \\
    0 & -\kappa_x & 0 }.
    \end{align}
    When $\kappa_z >1$, $\kappa_x$ becomes imaginary, and the wave becomes evanescent. In the main text we define $\kappa_z = \kappa$ and $\kappa_x = i \alpha$ to keep the variables real.
    
    The effect of the wave on freely falling test masses can be examined using the geodesic deviation equation
    \begin{equation}
    \frac{\partial^2 x^{\rho}}{\partial t^2} = -R^{\rho}{}_{0\nu0}(t)x^{\nu} = \frac{1}{2}\frac{\partial^2}{\partial t^2}(h^{\rho}{}_{\nu})^{\mathrm{TT}}(t)x^{\nu},
    \end{equation}
    where $x^{\nu}$ are vectors representing the proper positions of the test masses with respect to a fixed point. These equations have the solutions
    \begin{equation}
    x^{\rho}(t) = x_0^{\rho}+\Re[h^{\rho}{}_{\nu}(t)]x^{\nu}_0
    \end{equation}
    where $x_0^{\rho}$ is the initial condition (i.e. initial positions of the masses). If the waves are evanescent, $h^{\rho}{}_{\nu}^{\mathrm{TT}}$ has complex coefficients, and we can express the solution as a sum of its real and imaginary components
    \begin{equation*}
    x^{\rho}=x_0^{\rho}+\frac{1}{2}\Re(H^{\rho}{}_{\nu})x_0^{\nu}\cos(\omega{t}) + \frac{1}{2}\Im(H^{\rho}{}_{\nu})x_0^{\nu}\sin(\omega{t}).
    \end{equation*}
    Notice that for plane waves, $\Im(H^{\rho}{}_{\nu})^{\mathrm{TT}}=0$, and we recover the usual solution where test masses oscillate along a straight line with frequency $\omega$. The presence of the imaginary component which is off-phase to the real component means that the locus of test particles become ellipses as we asserted in the main text. 
    
    Also, masses which are initially at $x_0^i=(x_0,y_0,z_0)$ with $z_0=0$ or $y_0=0$ will move on a trajectory which is confined to a plane parallel to $\vec{k}'$. 
    To prove this we find the normal vector of this plane which is $\vec{a}\times\vec{b}$, where $a^i=\frac{1}{2}\Re(H^{j}{}_{i})^{\mathrm{TT}}x_0^{i}$ and $b^i=\frac{1}{2}\Im(H^{j}{}_{i})^{\mathrm{TT}}x_0^{i}$ and take a dot product with $\vec{k}'$. Without loss of generality we use $\vec{k}=k_0\qty(\ui\alpha,0,\kappa)$ and \cref{eq:evan} from the main text 
    \begin{equation}
        \vec{k}'\cdot\qty(\vec{a}\times\vec{b})=-(h^2_++h^2_\times)\kappa^2\alpha y_0z_0,
    \end{equation}
    note that this makes sense only if $\alpha\neq0$, because otherwise vector $\vec{a}\times\vec{b}$ does not exist.
%%%%%%%%%%%%%%%%%%%%%%%%%%%%%%%%%%%%%%%%%%%%%%%%%%%%%%%%%%%%%%%%%%%%%%%%%%%%%%%%%%%%%%%%%%%%%%%%%%%%%%%%%%%%%%%%%%%%%%%%%%%
\section{Transverse spin angular momentum\label{sec:tspin}}
    In this section we briefly discuss the transverse spin of evanescent gravitational waves. Transverse spin is a signature property of evanescent waves. It is extensively studied for electromagnetic waves~\cite{Bliokh2014,Bliokh2015c} but it has also been  recently discovered in acoustic waves~\cite{Bliokh2019}. 
    
    The most convenient way to study the spin of gravitational waves is using a \textit{Maxwellian form} of linearised gravity as proposed in \citet{Barnett2014}. This formalism introduces gravitational analogues of the electric and magnetic fields which can be expressed in terms of $h_{ij}$ in TT gauge as
    \begin{align}
        E_{ij}&=-{\partial}{_{t}}{h}{^{\mathrm{TT}}_{ij}}\\
        B_{ij}&={\epsilon}{_{jlm}}{\partial}{_{l}}{{h}}{^{{\mathrm{TT}}}_{im}}.
    \end{align}
    This treatment of the gravitational field is possible as long as we consider the weak field limit in a flat, Minkowski, background. One can then find the expression for spin angular momentum (SAM) density by calculating the Noether charge associated with rotations and isolating the spin part. The time averaged SAM is
    \begin{equation}\label{eq:spin}
        S^i
        =\dfrac{c^2}{64\pi G\ui\omega}\qty[E^\ast_{jm}E_{km}+c^2B^\ast_{jm}B_{km}]{\epsilon}{^{ijk}},
    \end{equation}
    noting that the complex conjugates come from the time average (here  $E_{ij}$ and $B_{ij}$ are considered to be phasors) with $c$ and $G$ kept explicit. In analogy with \citet{Bliokh2015c,Bliokh2019} we normalise the spin using the time-averaged energy density \begin{equation}
        W=\dfrac{c^2}{128\pi G}\qty[E^\ast_{ij}E^{ij}+c^2B^\ast_{ij}B^{ij}].
    \end{equation}
    Now one can examine the transverse spin in evanescent gravitational waves. Assuming an evanescent wave with $\vec{k}=\vec{k'}+\ui\vec{k}''$ and arbitrary polarisation as \cref{eq:evan} in the main text leads to 
    \begin{align}
        \vec{S}&=\frac{W}{\omega}\qty[2\frac{\sigma}{\kappa}\frac{\vec{k}'}{\qty|\vec{k}'|}+2\frac{\alpha}{\kappa}\frac{\vec{k}'\cp\vec{k}''}{\qty|\vec{k}'\cp\vec{k}''|}]\notag\\
        &=\frac{W}{\omega}\qty[2\sigma\frac{\omega}{c}\frac{\vec{k}'}{\qty|\vec{k}'|^2}+2\frac{\vec{k}'\cp\vec{k}''}{\qty|\vec{k}'|^2}],\label{eq:transversespin}
    \end{align}
    where $\sigma$ is a normalised third Stokes parameter, or \textit{helicity parameter}, which is defined as
    $$\sigma\equiv \frac{2\Im{h_+^\ast h_\times}}{\qty|h_+|^2+\qty|h_\times|^2}.$$
     As one might expect, it is manifest that for linearly polarised travelling plane waves, the longitudinal spin vanishes (as ${h}{^*_{ij}}={h}{_{ij}}$). \cref{eq:transversespin} shows that evanescent gravitational waves will acquire a transverse spin 
    \begin{equation}
        \dfrac{\omega \vec{S}_\perp}{W}=2\dfrac{\vec{k}'\cp\vec{k}''}{\qty|\vec{k}'|^2}
    \end{equation}
    which will be present for any polarisation and which is \textit{momentum locked} (if the direction of $\vec{k}'$ is reversed so will the direction of this spin). 
    
    Note the factor of $2$ in \cref{eq:transversespin}  arises when  Noether's theorem is applied to find \cref{eq:spin}
    \Cref{tab:spin} presents a comparison between gravitational, electromagnetic and acoustic waves. One can see that the transverse spin for an acoustic field also has the factor of two but in the case of an acoustic field this is due to the uneven contribution of the acoustic pressure $p$ and velocity $\vec{v}$ (these fields play the role of $\vec{E}$ and $\vec{H}$ fields in acoustics) to the SAM \cite{Bliokh2019}.
     \begin{table*}[!ht]
        \centering
        \caption[Comparison of linearised gravity, electromagnetism and acoustics]{The extension of (Table I) published in~\citet{Bliokh2019}. This table compares gravity, electromagnetism and acoustics. Here ${\bar{h}}{_{\mu\nu}}$ is a trace-reversed metric perturbation and ${h}{^{\mathrm{TT}}_{ij}}$ is its transverse traceless part. $A^\nu$ is an electromagnetic four-potential and $\vec{A}^\perp$ is its transverse part. For the gravitational field we define $\varepsilon=\varepsilon_0\varepsilon_r$ and $\mu=\mu_0\mu_r$ (in analogy to electromagnetism), where $1/\varepsilon_0=c^2\mu_0=32\pi Gc^{-2}$. Quantities $W$ and $\vec{S}$ are averaged over a period.}
        \begingroup % keep the change local
    \setlength{\tabcolsep}{5pt}
    \renewcommand{\arraystretch}{2.5}
        \begin{tabular}{ | l | c | c | c | }
        \cline{2-4}
        \multicolumn{1}{c|}{}&Linearised Gravity&Electromagnetism&Acoustics 
         \\%[1em]
        \hline
        potentials&
        ${\bar{h}}{^{\mu\nu}}$&
        ${A}{^{\nu}}$&
        $\varphi$
         \\%[1em]
        \hline
        wave equation& 
        {$\Box\,{{\bar{h}}{_{\mu\nu}}}=0$}&
        {$\Box\,{{A}{_{\nu}}}=0$}&
        {$\Box\,{\varphi}=0$} 
        \\%[1em]
        \hline
        gauge condition& {$\partial_\mu{{\bar{h}}{^\mu_\nu}}=0$}&
        {$\partial_\nu{{A}{^{\nu}}}=0$}& {$-$}
        \\%[1em]
        \hline
        fields
        &
        {
        \begin{minipage}[c]{3.2cm}\vspace{-0.5\baselineskip}
        \begin{align*}
            E_{ij}&=-{\partial}{_{t}}{h}{^{\mathrm{TT}}_{ij}}
            \\
        \mu H_{ij}&={\epsilon}{_{jlm}}{\partial}{_{l}}(h^{{\mathrm{TT}}})_{i}{}^{m}
        \end{align*}\vspace{-0.6\baselineskip}
        \end{minipage}}
        &
         {
        \begin{minipage}{3cm}\vspace{-0.5\baselineskip}
        \begin{align*}
            \vec{E}&=-{\partial}{_{t}}\vec{A}^\perp
            \\
            \mu\vec{H}&=\curl{\vec{A}^\perp}
        \end{align*}\vspace{-0.6\baselineskip}
        \end{minipage}}
        &{
        \begin{minipage}{3cm}\vspace{-0.5\baselineskip}
        \begin{align*}
            p&=\rho~{\partial}{_{t}}\varphi
            \\
            \vec{v}&=\grad{\varphi}
        \end{align*}\vspace{-0.6\baselineskip}
        \end{minipage} 
        }
        \\%[1em]
        \hline
        constraints&${\partial}{^{i}}{E}{_{ij}}={\partial}{^{i}}{H}{_{ij}}=0$&
        $\div{\vec{E}}=\div{\vec{H}}=0$
        &$\curl{\vec{v}}=\vec{0}$ 
        \\%[1em]
        \hline
        medium parameters
        &
        {$\varepsilon$, $\mu$} &
        {$\varepsilon$, $\mu$}&
        {$\rho$, $\beta$}
        \\%[1em]
        \hline
        energy density $W$
        &
        {$\dfrac{1}{4}\qty(\varepsilon E^\ast_{ij}E^{ij}+\mu H^\ast_{ij}H^{ij})$}&
        {$\dfrac{1}{4}\qty(\varepsilon\qty|\vec{E}|^2+\mu\qty|\vec{H}|^2)$}& 
        {$\dfrac{1}{4}\qty(\beta\qty|p|^2+\rho\qty|\vec{v}|^2)$}
        \\%[1em]
        \hline 
        SAM density $\vec{S}$
        &
        {$\dfrac{\epsilon^{ijk}}{2\ui\omega}\qty\Big[\varepsilon E^\ast_{jm}E_{km}+\mu H^\ast_{jm}H_{km}]$}
        &
        {$\dfrac{1}{4\ui\omega}\qty\Big[\varepsilon\vec{E}^\ast\cp\vec{E} +\mu\vec{H^\ast}\cp\vec{H}]$}
        & 
        {$\dfrac{1}{2\ui\omega}\qty\Big[\rho\vec{v}^\ast\cp\vec{v}]$}
        \\%[1em]
        \hline
        transverse spin density& 
        {$\dfrac{\omega \vec{S}_\perp}{W}=2\dfrac{\vec{k}'\cp\vec{k}''}{\qty|\vec{k}'|^2}$}&
        {$\dfrac{\omega \vec{S}_\perp}{W}=\dfrac{\vec{k}'\cp\vec{k}''}{\qty|\vec{k}'|^2}$}& 
        {$\dfrac{\omega \vec{S}_\perp}{W}=2\dfrac{\vec{k}'\cp\vec{k}''}{\qty|\vec{k}'|^2}$}
        \\%[1em]
        \hline
        longitudinal spin density& 
        {$\dfrac{\omega \vec{S}_\parallel}{W}=2\dfrac{\sigma}{\kappa}\dfrac{\vec{k}'}{\qty|\vec{k}'|}$}&
        {$\dfrac{\omega \vec{S}_\parallel}{W}=\dfrac{\sigma}{\kappa}\dfrac{\vec{k}'}{\qty|\vec{k}'|}$}& 
        {$\dfrac{\omega \vec{S}_\parallel}{W}=0$}
        \\%[1em]
        \hline
        \end{tabular}
        \endgroup
        \label{tab:spin}
    \end{table*}
%%%%%%%%%%%%%%%%%%%%%%%%%%%%%%%%%%%%%%%%%%%%%%%%%%%%%%%%%%%%%%%%%%%%%%%%%%%%%%%%%%%%%%%%%%%%%%%%%%%%%%%%%%%%%%%%%%%%%%%%%%%
\section{Local description as a linear combination of travelling waves}\label{sec:localdes}
    In analogy with electromagnetism, it is possible to locally describe the polarisation of an evanescent wave as that of two interfering plane waves travelling in mutually orthogonal directions. Without loss of generality, for any evanescent wave it is possible to find a frame where 
    \begin{align}
    e_{ij}^{+}(\vec{k}'+\ui\vec{k}'') &= 
    \Pmqty{ 
    \kappa^2 & 0 & -\ui\alpha\kappa \\
    0 & -1 & 0 \\
    -\ui\alpha\kappa& 0 & -\alpha^2 \\
    }~,~\\
    e_{ij}^{\times}(\vec{k}'+\ui\vec{k}'') &= 
    \Pmqty{ 
    0 & \kappa & 0 \\
    \kappa & 0 & -\ui\alpha \\
    0 & -\ui\alpha & 0 \\
    },
    \end{align}
    with $\kappa^2=1+\alpha^2$ being the null condition. Now one can see that 
    \begin{equation*}
    \Pmqty{ 
    \kappa^2 & 0 & -\ui\alpha\kappa \\
    0 & -1 & 0 \\
    -\ui\alpha\kappa& 0 & -\alpha^2 \\
    }=
    \Pmqty{ 
    1& 0 & 0 \\
    0 & -1 & 0 \\
    0& 0 & 0 \\
    }+
    \Pmqty{  
    \alpha^2 & 0 & -\ui\alpha\kappa \\
    0 & 0 & 0 \\
    -\ui\alpha\kappa& 0 & -\alpha^2 \\
    },
    \end{equation*}
    which is a linear combination of a plus polarised wave in the direction of $\vec{k}'$ and an elliptically polarised wave in the direction of $\vec{k}'\times\vec{k}''$ 
    \begin{align*}
    e_{ij}^{+}(\vec{k}'&+\ui\vec{k}'') =\\ &e_{ij}^{+}(\vec{k}')+\left(\alpha^2e_{ij}^{+}(\vec{k}'\times\vec{k}'')-\ui\alpha\kappa e_{ij}^{\times}(\vec{k}'\times\vec{k}'')\right).
    \end{align*}
    Similarly for the cross polarisation 
    \begin{equation*}
    \Pmqty{  
    0 & \kappa & 0 \\
    \kappa & 0 & -\ui\alpha \\
    0 & -\ui\alpha & 0 \\
    }=\kappa
    \Pmqty{ 
    0& 1 & 0 \\
    1 & 0 & 0 \\
    0& 0 & 0 \\
    }-\ui\alpha
    \Pmqty{ 
    0 & 0 & 0 \\
    0 & 0 & 1 \\
    0 & 1 & 0 \\
    },
    \end{equation*}
    which is a linear combination of a cross polarised wave in the direction of $\vec{k}'$ and an out-of-phase cross polarised wave in the direction of $\vec{k}''$
    \begin{equation*}
    e_{ij}^{\times}(\vec{k}'+\ui\vec{k}'') = \kappa e_{ij}^{\times}(\vec{k}')-\ui\alpha e_{ij}^{\times}(\vec{k}'').
    \end{equation*}
%%%%%%%%%%%%%%%%%%%%%%%%%%%%%%%%%%%%%%%%%%%%%%%%%%%%%%%%%%%%%%%%%%%%%%%%%%%%%%%%%%%%%%%%%%%%%%%%%%%%%%%%%%%%%%%%%%%%%%%%%%%
\section{Angular spectrum representation}
    To show that there are evanescent components near a sub-wavelength source of gravitational radiation, we will use the angular spectrum representation. This is a standard tool for studying wave-fields in homogeneous media and is widely used in nanophotonics to study scattering, beam propagation, focusing, holography, and many other phenomena \cite{Mandel1995}. The main idea is that, in general, solutions to the wave equation are not momentum eigenmodes with a well-defined wave-vector $\vec{k}$ as in \cref{eq:tensor}, but rather a distribution of them constituting a continuous spectrum. The generalisation of this representation to higher ranked tensors such as vectors and rank-2 tensor gravitational waves can be made, as shown below. Suppose we know a field ${h}{_{ij}}(\vec{r},t)$ at any point. We may assume it is time-harmonic without loss of generality as we can always perform a temporal Fourier transform. We can consider this field in a plane $z=\text{constant}$ which is transverse to an arbitrary $z$-direction. In this plane, we can write the field as a 2D inverse spatial Fourier transform \cite{Mandel1995}:
     \begin{align*}
        {h}{_{ij}}\qty(x,y,z,t)=&\iint\limits_{-\infty}^{\phantom{--}\infty}{{\tilde{h}}{_{ij}}(k_x,k_y;z)}\ue^{\ui\qty(k_xx+k_yy-\omega t)}\dd{k_x}\dd{k_y}.
     \end{align*}
        Since the wave satisfies the wave equation and hence the null condition \cref{eq:null} it is possible to ``propagate'' the field from the source plane to any other plane with different $z=\text{constant}$ via a simple multiplicative transfer function \cite{Mandel1995}:
    \begin{align*}
        \tilde{h}{_{ij}}(k_x,k_y;z)&={\tilde{h}_{ij}^{(+)} \left(k_x,k_y;0\right) }\ue^{+\ui k_zz}\\&+{\tilde{h}_{ij}^{(-)} \left(k_x,k_y;0\right) }\ue^{- \ui k_zz}.
    \end{align*}
        where the two terms $(+)$ and $(-)$, not to be confused with the $+$ and $\cross$ modes, account for the two possible signs of $\pm k_z =\pm (k_0^2-k_x^2-k_y^2)^{1/2}$. When the fields originate from a localised source at $z=0$, only the sign of $k_z$ propagating away from the source needs to be considered; hence we use the plus representation for $z>0$ and the minus representation for $z<0$. This means that a complete knowledge of the fields in the entirety of space can be gained from one single plane. A key simplification can be made by realising that the integrand becomes ${\tilde{h}_{ij}^{(\pm)}(k_x,k_y;0)}\ue^{\ui\qty(k_xx+k_yy\pm k_zz-\omega t)}$, which has exactly the same form as \cref{eq:tensor} in the main text, and therefore, for each value of transverse momentum $(k_x,k_y)$, it must fulfil \cref{eq:trans,eq:null}, and can thus be reduced to a superposition of only two tensor modes as in \cref{eq:evan}:
    \begin{equation}
        \tilde{h}^{(\pm)}_{ij}(k_x,k_y;0)=h_+^{(\pm)}{e}{^+_{ij}}+ h_{\times}^{(\pm)}{e}{^{\times}_{ij}},\label{eq:decompostap}
    \end{equation}
        where ${e}{^{+}_{ij}}(k_x,k_y,\pm k_z)$ and ${e}{^{\times}_{ij}}(k_x,k_y, \pm k_z)$ are the same complex basis tensors introduced in \cref{eq:evan}, but generalised for arbitrary directions (see \cref{sec:A}). Therefore, a pair of scalar complex-valued angular spectra $h_+^{(\pm)}(k_x,k_y)$ and $h_\times^{(\pm)}(k_x,k_y)$ completely describe the source in momentum space, and hence in all of real space via the spectral representation. These two spectra include all information of the amplitude and phase of the two polarisation modes of the propagating far field plane waves in every direction (the momentum representation in the region $k_x^2+k_y^2\leq k_0^2$, corresponding to real $k_z$) and also tell us about all the evanescent near fields components (in the region $k_x^2+k_y^2>k_0^2$, corresponding to imaginary $k_z$) whose amplitude decays exponentially as $|z|$ is increased.
%%%%%%%%%%%%%%%%%%%%%%%%%%%%%%%%%%%%%%%%%%%%%%%%%%%%%%%%%%%%%%%%%%%%%%%%%%%%%%%%%%%%%%%%%%%%%%%%%%%%%%%%%%%%%%%%%%%%%%%%%%%
\section{General solution of linearised Einstein equation for binary system}
\begin{figure}[t!] 
         \centering
         \includegraphics{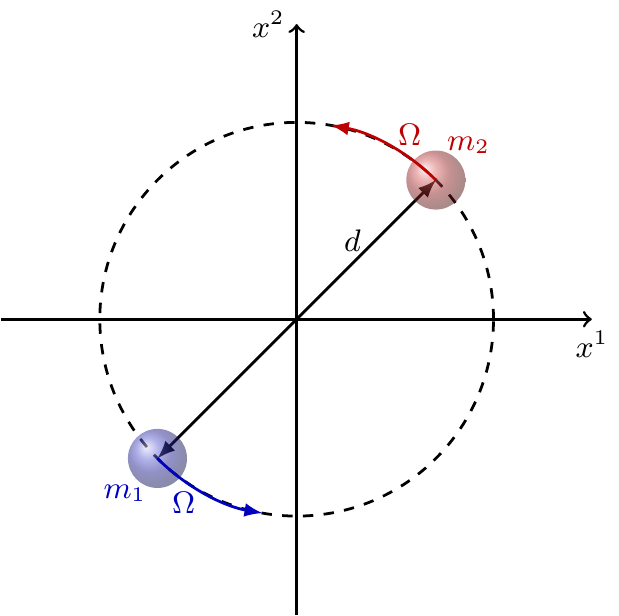}
          \caption[Compact binary system scheme]{Compact binary system modelled as two point masses, $m_1=m_2=M$ orbiting around common centre of mass with angular frequency $\Omega$.}\label{fig:CBS}
    \end{figure}

    Consider a compact binary system, modelled as two point masses $m_1=m_2=M$, orbiting around a common centre of mass with angular frequency $\Omega$ (see \cref{fig:CBS}). We can describe locations in space-time using the coordinates $x^\mu=(ct,x^i)$ of an observer who is located outside the source in a region where the gravitational field is linear $g_{\mu\nu}=\eta_{\mu\nu}+h_{\mu\nu}$, where $\eta_{\mu\nu}$ is the Minkowski metric used to raise and lower Greek indices because it is the background metric in this setting. Each of the masses moves along its world-line which can be represented by a parametric curve $X^\mu_s(\tau_s)$, where $s\in\{1,2\}$ labels the mass and $\tau_s$ is the proper time of the respective mass. We assume trajectories of these masses in the coordinates of the outside observer to be perfect circles
    \begin{equation}\label{eq:obstraj}
        X_s^\mu(t) = 
      \pmqty{ct,(-1)^s\frac{d}{2}\cos(\Omega t),(-1)^s\frac{d}{2}\sin(\Omega t),0}^\intercal~
    \end{equation}
    The matter action for this system is a sum of the actions for two point masses 
    \begin{equation}\label{eq:action1}
        \mathcal{S}[g,X_1,X_2]=-\sum_{s=1}^2\int Mc\sqrt{g_{\mu\nu}\displaystyle\dot{X}^\mu_s\dot{X}^\nu_s}\dd{\tau_s}.
    \end{equation}
    Here we used the fact that world-lines are sub-manifolds of space-time and the metric can be pulled back onto each of the world-lines 
    \begin{equation}
        \dd{s}^2=g_{\mu\nu}\dd{x^\mu}\dd{x^\nu}\mapsto g_{\mu\nu}\dot{X}^\mu_s\dot{X}^\nu_s\dd{\tau}_s^2,
    \end{equation}
    where dot denotes derivative with respect to the proper time, hence $\dot{X}^\mu_s$ is a four-velocity. The mass can be written as $M=\int\dd[4]{x}M\delta^{(4)}(x^\mu-X^\mu_s(\tau_s))=\int\dd[4]{x}M\delta(t-\tau_s)\delta^{(3)}(x^i-X^i_s(\tau_s))$. This, together with the fact that $\dd{\tau_s}=\dv*{\tau_s}{t}\dd{t}$, allows us to write \cref{eq:action1} as
    \begin{align*}
        % \mathcal{S}&=-\sum_{s=1}^2Mc\int\dd[4]{x}\int\dv{\tau_s}{t}\delta^{(3)}(x^i-X_s^i(\tau))\sqrt{g_{\mu\nu}\displaystyle\dot{X}^\mu_s\dot{X}^\nu_s}\dd{t}\delta(t-\tau_s) \notag\\
        \mathcal{S}&=-\sum_{s=1}^2Mc\int\dd[4]{x}\dv{\tau_s}{t}\delta^{(3)}(x^i-X^i_s(t))  \sqrt{g_{\mu\nu}\displaystyle\dot{X}^\mu_s\dot{X}^\nu_s}
    \end{align*}
    Using this action one can find the Hilbert stress-energy tensor $T_{\mu\nu}$ using the definition
    \begin{equation}
        T_{\mu \nu}=\frac{-2}{\sqrt{-\det g}} \fdv{\mathcal{S}}{g^{\mu \nu}}.
    \end{equation}
    By plugging in the action and recognising that the magnitude of the four-velocity is always $c$ and $\det \eta_{\mu\nu}=-1$,
    \begin{equation}
        T_{\mu \nu}=\sum_{s=1}^2M\dv{\tau_s}{t}{\dot{X}_\mu^s\dot{X}_\nu^s}\delta^{(3)}(x^i-X^i_s(t)).
    \end{equation}  
    Given \cref{eq:obstraj} it is more convenient to write this as 
    \begin{equation}\label{eq:Tmn}
        T_{\mu \nu}(t,x^i)=\sum_{s=1}^2M\gamma(v){v_\mu^sv_\nu^s}\delta^{(3)}(x^i-X^i_s(t)),
    \end{equation}  
    where $v^\mu_s=\dv*{X^\mu_s}{t}$ is the velocity observed by the observer and $\gamma(v)=\dv*{t}{\tau_s}$ is the usual Lorentz factor for the orbital speed $v=\sqrt {v^iv_i}=\Omega d/2$.
    
    \begin{figure*}[!t]
        \includegraphics[width=.8\textwidth]{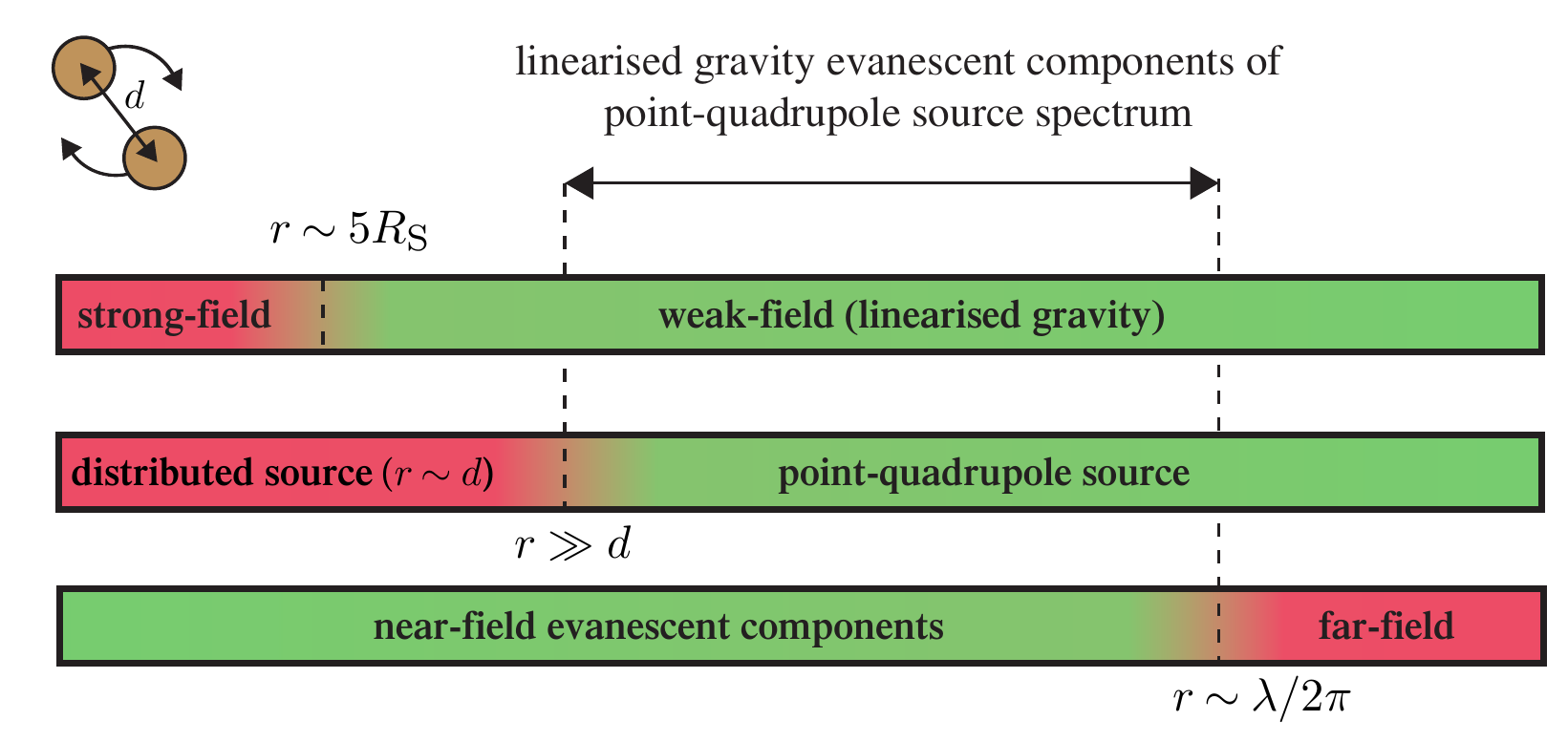}
        \caption{Regions of space around a source of gravitational waves. The validity of the different approximations changes gradually, but we indicate the approximate distances where they are generally considered valid. Adapted from \citet{Thorne1983}}
        \label{fig:zones}
    \end{figure*}    
    
    Now we want to solve the linearised Einstein equation,
    \begin{equation}\label{eq:linei}
        \bar{h}_{\mu\nu}=-\frac{16\pi G}{c^4}T_{\mu \nu},
    \end{equation}
    using the retarded Green's function \cite{Maggiore2007}
    \begin{align*}\label{eq:retardA}
        {\bar{h}}_{\mu\nu}\left(t,x^i\right)=\frac{4G}{c^4}\int\dd[4]{{x} '} &\frac {{T}_{\mu\nu}\left(t',x'^i \right)}{\left|\vec{x} -\vec{x} '\right|}\\&\delta\qty\big(ct'-ct+\left|\vec{x} -\vec{x} '\right|).
    \end{align*}
    At this point, people conventionally assume the far-field limit to simplify this integral, but since we are interested in the near-field behaviour we shall take no such limit. Instead, we consider the full analytical solution for this problem.
    For our compact binary with energy momentum tensor \cref{eq:Tmn} we will have
    \begin{align*}
    {\bar{h}}_{\mu\nu}&=\frac{4GM\gamma(v)}{c^4}\int\dd[4]{{x} '}\sum_{s=1}^2\frac {{v_\mu^s(t')v_\nu^s(t')}}{\left|\vec{x} -\vec{x} '\right|}\\
    &\phantom{=}\delta^{(3)}\qty\big({x}'^i-{X}^i_s(t')) \delta\qty\big(ct'-ct+{\left|\vec{x} -\vec{x} '\right|}),
    \end{align*}
    note that $\gamma(v)$ is the same for both masses since $v=\Omega d/2$ is a constant of the system. Also, $\dd[4]{{x}'}=\dd{ct'}\dd[3]{\vec{x}'}$ so we can perform integration over the spatial coordinates separately
    \begin{align*}
    {\bar{h}}_{\mu\nu}&=\frac{4GM\gamma(v)}{c^4}\int c\dd{t'}\sum_{s=1}^2\frac {{v_\mu^s(t')v_\nu^s(t')}}{\left|\vec{x} -\vec{X}_s(t')\right|}\\
    &\phantom{=} \delta\qty\big(ct'-ct+{\left|\vec{x} -\vec{X}_s(t')\right|}).
    \end{align*}
    We cannot now simply use the delta function to integrate since the argument of this delta function is not a linear function of $ct'$. This can be bypassed using the following identity 
    \begin{equation}
    \delta(g(x))=\frac{\delta\left(x-x_{0}\right)}{\left|g^{\prime}\left(x_{0}\right)\right|}
    \end{equation}
    which holds provided that $g$ is a continuously differentiable function with a real root at $x_{0}$ and with nowhere vanishing derivative $g'$. This allows us to convert
    \begin{equation*}
    \delta\qty\big(ct'-ct+{\left|\vec{x} -\vec{X}_s(t')\right|})=\frac{\delta\qty\big(ct'-ct+{\left|\vec{x} -\vec{X}_s(t_s)\right|})}{\left|1-\frac{\vec{x} -\vec{X}_s(t_s)}{{\left|\vec{x} -\vec{X}_s(t_s)\right|}}\vdot\frac{\vec{v}_s(t_s)}{c}\right|}
    \end{equation*}
    where $t_s$ is a retarded time at which a signal from a source $s$ is received by the observer, defined by the implicit relation
    \begin{equation}
    t_s=t-\frac{\left|\vec{x} -\vec{X}_s(t_s)\right|}{c}.
    \end{equation}
    Now we can use this new delta function which only linearly depends on $t'$ to finally arrive at the solution
    \begin{align*}
    {\bar{h}}_{\mu\nu}&=\frac{4GM\gamma(v)}{c^4}\sum_{s=1}^2\frac {{v_\mu^s(t_s)v_\nu^s(t_s)}}{\left|\vec{x} -\vec{X}_s(t_s)\right|-\qty[\vec{x} -\vec{X}_s(t_s)]\vdot\frac{\vec{v}_s(t_s)}{c}}.
    \end{align*}
    Until this point we have considered a rather general solution, only assuming that we are in a weak-field zone. According to \citet{Thorne1983}, the weak-field zone starts at five Schwarzschild radii ($R_\mathrm{S}=2GM/c^2$), as depicted in \cref{fig:zones}.
    
%%%%%%%%%%%%%%%%%%%%%%%%%%%%%%%%%%%%%%%%%%%%%%%%%%%%%%%%%%%%%%%%%%%%%%%%%%%%%%%%%%%%%%%%%%%%%%%%%%%%%%%%%%%%%%%%%%%%%%%%%%%
\section{Proof of quadrupolar radiation being dominant in the near field}
    We argue in the main text that even in the near field zone of a sub-wavelength source (a source with wavelength $\lambda\gg d$) it is possible to use the quadrupole formula. Note that such source has to be slow-moving/non-relativistic since $v=\Omega d/2=\omega d/4$, where $\omega=2\Omega$ is the characteristic frequency of gravitational waves produced by this source, and $v/c=\omega d/(4c)=\pi d/(2\lambda)$ so $\lambda\gg d$ implies that $c\gg v$. If we now look at 
    \begin{align*}
    {\bar{h}}_{\mu\nu}&=\frac{4GM\gamma(v)}{c^4}\sum_{s=1}^2\frac {{v_\mu^s(t_s)v_\nu^s(t_s)}}{\left|\vec{x} -\vec{X}_s(t_s)\right|-\qty[\vec{x} -\vec{X}_s(t_s)]\vdot\frac{\vec{v}_s(t_s)}{c}}.
    \end{align*}
    we can find that 
    \begin{align}\label{eq:nearquadrupole}
    {\bar{h}}_{\mu\nu}&=\frac{4GM}{c^4}\sum_{s=1}^2\frac {{v_\mu^s(t_s)v_\nu^s(t_s)}}{\left|\vec{x} -\vec{X}_s(t_s)\right|}\qty[1+\order{\frac{v}{c}}], 
    \end{align}
    where we have used the expansions
    \begin{align*}
        \gamma(v)&=1+\frac{v}{c}+\order{\frac{v^4}{c^4}}\\
        \intertext{and}
        \frac{1}{1-\frac{\vec{v}\vdot\vu{n}}{c}}&=1+\frac{\vec{v}\vdot\vu{n}}{c}+\frac{(\vec{v}\vdot\vu{n})^2}{c^2}+\order{\frac{(\vec{v}\vdot\vu{n})^3}{c^3}},
    \intertext{which together with the Cauchy--Bunyakovsky--Schwarz inequality $|\langle\mathbf{u}, \mathbf{v}\rangle| \leq\|\mathbf{u}\|\|\mathbf{v}\|$ yields}
    \frac{\gamma(v)}{1-\frac{\vec{v}\vdot\vu{n}}{c}}&\leq1+\frac{v}{c}+\frac{3}{2}\frac{v^2}{c^2}+\order{\frac{v^3}{c^3}}.
    \end{align*}
    One can see that \cref{eq:nearquadrupole} starts to look nearly like the quadrupolar formula, we only now have to take care of the retarded times $t_s$ and retarded positions ${\left|\vec{x} -\vec{X}_s(t_s)\right|}$. That can be done by carefully placing the observer in the near field, but far enough to the source so that the point-quadrupole approximation holds (see figure \cref{fig:zones}). In other words we choose $\abs{\vec{x}}=r$ to be the same order of magnitude as $\lambda$ so that we can use $d/r\sim d/\lambda\ll1$. Note that this allows us to take $d\ll r<\lambda/2\pi$, which means we can have the observer in a zone where we can observe evanescent fields ($r<\lambda/2\pi$). Finally we get 
    \begin{align*}
    {\bar{h}}_{\mu\nu}&=\frac{4GM}{c^4}\sum_{s=1}^2\frac {{v_\mu^s(t_r)v_\nu^s(t_r)}}{r}\qty[1+\order{\frac{v}{c}}+\order{\frac{d}{r}}], 
    \end{align*}
    where $t_r=t-r/c$ and we have used that 
    \begin{equation*}
        {\left|\vec{x} -\vec{X}_s(t_s)\right|}=r\qty[1+\order{\frac{d}{r}}].
    \end{equation*}
    In the induction zone (for a sub-wavelength source) the formula simplifies to the well known quadrupole formula, since $-v_i^1(t)=v_i^2(t)=v_i(t)$
    \begin{align*}
    {\bar{h}}_{ij}&=\frac{8GM}{c^4}\frac {{v_i(t_r)v_j(t_r)}}{r}
    \end{align*}
    Therefore in this limit we can consider our source to be a point gravitational quadrupole as we did in the main text.
%%%%%%%%%%%%%%%%%%%%%%%%%%%%%%%%%%%%%%%%%%%%%%%%%%%%%%%%%%%%%%%%%%%%%%%%%%%%%%%%%%%%%%%%%%%%%%%%%%%%%%%%%%%%%%%%%%%%%%%%%%%
    \begin{figure*}[!t]
        \includegraphics{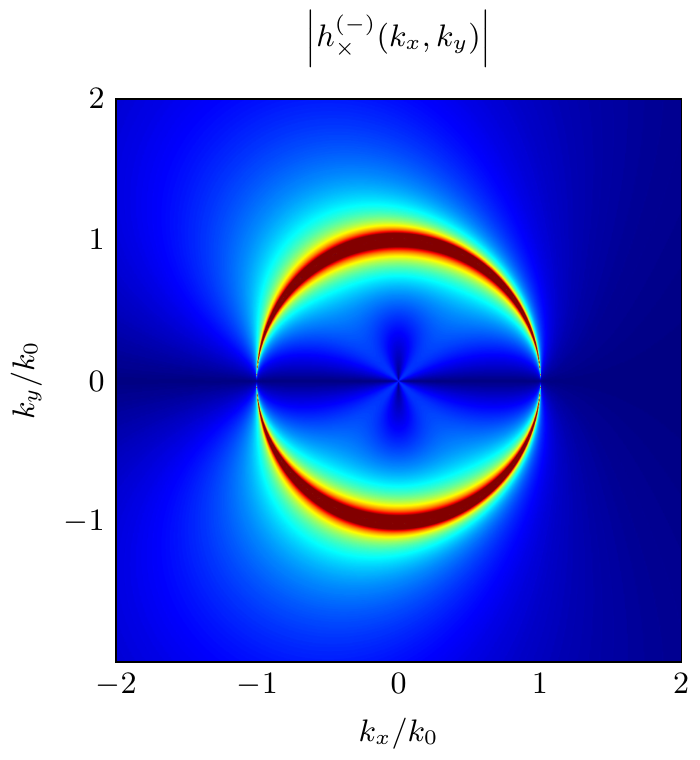}
        \includegraphics{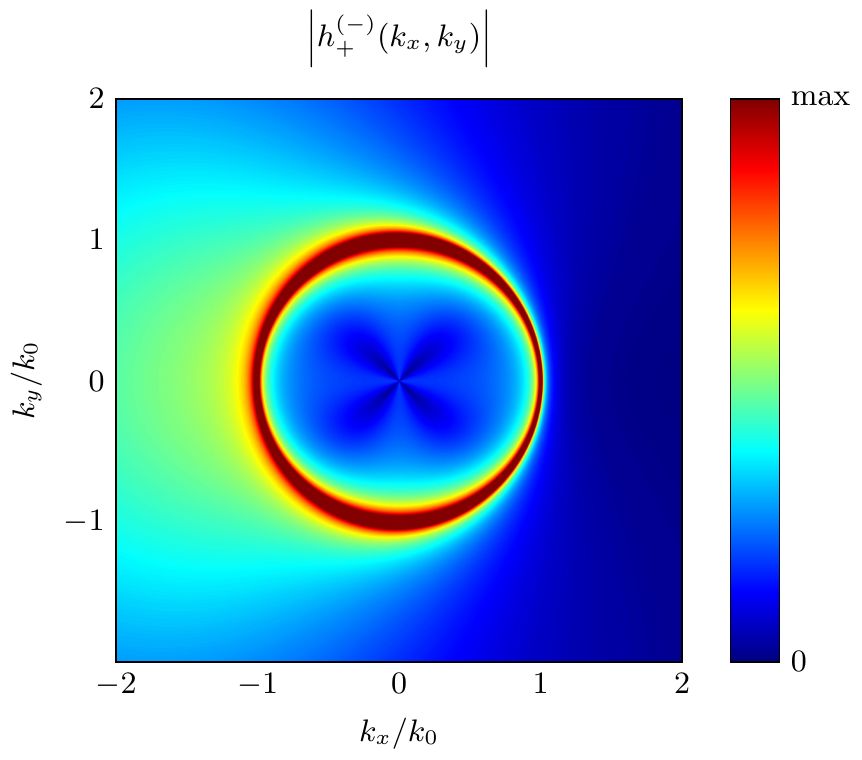}
        \caption{Plot of complex amplitudes of the two spectra $|h_+^{(-)}|$ and $|h_\times^{(-)}|$  after propagating them to a plane $z = -\lambda/2\pi$ via the transfer function $\ue^{- \ui k_z z}$.}\label{fig:xza}
    \end{figure*}
    
\section{Evanescent gravitational waves near a quadrupolar source}\label{sec:genesis}
    A system with two equal masses $M$ orbiting each other at a distance $d$ in a stable circular orbit with frequency $\Omega$ around a common centre of mass on the $x$-$z$ plane (\cref{fig:xz}) is given by
    \begin{align}
    {\vec{x}}_s(t) &= 
     \frac{d}{2} \Pmqty{\cos(\Omega t),&0,&\sin(\Omega t)}^\intercal~,\quad \notag\\\rho(t,{\vec{x}}) &= M\left[ \delta^3({\vec{x}}-{\vec{x}}_s(t)) + \delta^3({\vec{x}}+{\vec{x}}_s(t)) \right].
     \end{align}
    Assuming that
    \begin{itemize}
        \item the field is weak $|h|\ll1$,
        \item point masses are in stable circular orbit,
        \item the speed of the source is not relativistic $v\ll c$, which directly implies that the source is sub-wavelength ${\vec{k}}\cdot{\vec{d}}\ll 1 \Leftrightarrow 2d\ll \lambda/2\pi$,
        \item  The observer is not too close to the source (at distance comparable to $2d$) which does not necessarily mean that the observer is in the far field thanks to the previous assumption. The observer can be at distance $r$ for which $2d\ll r \ll \lambda$.\end{itemize} 
    
    This system has the well-known solution
    \begin{equation}
    h_{ij}(t,r) = \frac{2G}{c^4}\frac{1}{r}\frac{d^2}{dt^2}Q_{ij}(t-r/c)~,~\label{eqn:quadrupole}
    \end{equation}
    where $Q_{ij}$ is the quadrupole moment of mass density $\rho$ 
    \begin{equation}
    Q_{ij} = \iiint\limits_{\mathbb{R}^3} \rho(t,{\vec{x}})x_ix_j d^3x.
    \end{equation}
    
    The solution  of \cref{eqn:quadrupole}, in frequency space is given by
    \begin{equation}
    h_{ij}(\omega, r) = \frac{G \omega^2}{4c^4} q_{ij} \frac{\ue^{\ui\omega r/c}}{r} \label{eqn:hijspherical},
    \end{equation}
    where 
    \begin{equation}
    q_{ij}=\frac{Md^2}{2}
     \left(\begin{array}{ccc}
     -1 & 0 & -\ui \\
     0 & 0 & 0 \\
     -\ui & 0 & 1 \\
        \end{array}
        \right).
     \end{equation}
    
    \cref{eqn:hijspherical} describes a spherical wave, with origin at $r=0$ or ${\vec{x}}=0$. At $r\gg \lambda$, the wavefront is asymptotic to a plane wave locally. However, in the near zone, this is not true. Our goal hence is to find a decomposition of a spherical wave into a spectrum of plane waves, labelled by $(k_x,k_y,k_z)$.  This angular spectrum representation is a standard problem in the study of reflection and refraction of spherical waves. This decomposition can be performed by first choosing a special axis (we choose $z$), and then performing a 2-D Fourier transform on the plane (we choose $(x,y)$)  as follows:
    \begin{equation}
    \frac{\ue^{\ui\omega r/c}}{r} = \frac{\ui}{2\pi}\iint\limits^{\phantom{-}\infty}_{-\infty}\frac{1}{k_z}\ue^{\ui k_xx+\ui k_yy}\ue^{\pm \ui k_z z}~\dd{k_x}\dd{k_y}, \label{eqn:weylrepresentation}
    \end{equation}
    where we choose the plus representation for $z>0$ and the minus representation for $z<0$. This representation is often known as the Weyl identity and its derivation can be found in \citet{Mandel1995}. By simple substitution of \cref{eqn:weylrepresentation} into \cref{eqn:hijspherical} we can rewrite \cref{eqn:hijspherical} as the following spectrum of plane waves
    \begin{equation}
    h_{ij}(\omega,x,y,z) = \iint\limits^{\phantom{-}\infty}_{-\infty} \tilde{h}_{ij}^{(\pm)}(\omega, k_x,k_y,z)\ue^{\ui k_xx+\ui k_yy}~\dd{k_x}\dd{k_y},
    \end{equation}
    with 
    \begin{equation}
    \tilde{h}_{ij}^{(\pm)}(\omega,k_x,k_y,z) = \frac{G}{8\pi c^4} \frac{\ui\omega^2}{k_z}q_{ij}\ue^{\pm \ui k_z z}~\label{eqn:spectrumhij}.
    \end{equation}
    Note that this decomposition, and the resulting spectrum \cref{eqn:spectrumhij}, breaks the spherical symmetry of \cref{eqn:hijspherical}. The two mode basis tensors obey the relations $e^{ij}_\times e_{ij}^\times = e^{ij}_+ e_{ij}^+ = 2$ and $e^{ij}_\times e_{ij}^+ = e^{ij}_+ e_{ij}^\times = 0$, so the amplitudes $h_+$ and $h_-$ of the modes can be obtained by projection, as:
    \begin{align}
    h_+^{(\pm)} &=\frac{1}{2} \tilde{h}_{ij}^{(\pm)}(\omega,k_x,k_y,z)e^{ij}_+(\omega,k_x,k_y,\pm k_z),\\
    ~h_\times^{(\pm)} &= \frac{1}{2} \tilde{h}_{ij}^{(\pm)}(\omega,k_x,k_y,z)e^{ij}_\times(\omega,k_x,k_y,\pm k_z).
    \end{align}
    The key point here is that, for a chosen $z$, the spectrum has a very different behaviour in different regions of the $(k_x,k_y)$ plane, depending on the sign of $k_z^2\equiv \omega^2/c^2-k_x^2-k_y^2$. If $k_z^2<0$, then the plane wave is evanescent. This occurs outside the circle of radius $\omega/c$ in the $(k_x,k_y)$ plane. Notice that in this case, the $\ue^{\pm \ui k_z z}$ term on the RHS of \cref{eqn:spectrumhij} suppresses the contribution of evanescent waves as $z\rightarrow \infty$. In \cref{fig:xza} we show that, unsurprisingly, the spectrum contains standard plane waves $k_z^2>0$ which describe the far-field radiation diagram of a quadrupole source. However, the spectrum also exhibits support for the $k_z^2<0$ evanescent components, which in fact dominate in amplitude when sufficiently close to the source.
    \begin{table}[!ht]
    \caption{The near-field zones of present and future detectors, calculated as $d<\lambda/2\pi=c/2\pi f$.}
    \begin{tabular}{|l|l|l|}
    \hline
     & Frequency [Hz] & Near zone [A.U.]   \\ \hline
     LIGO &  $10^{3}$ & $10^{-7}$     \\ \hline
     LISA & $10^{-2}$ & $10^{-2}$    \\ \hline
     ETA & $10^{2} $ &  $10^{-6}$ \\
     \hline
    \end{tabular}
    \label{table:nearzone}
    \end{table}
    
    As can be seen in Table \ref{table:nearzone}, the near zones for the frequencies of present and near future GW directors are very close to the source, and hence direct detection of EGW from binaries is not very likely.

\end{document}